\documentclass[amsmath,amssymb,aps,prd,11pt,tightenlines,superscriptaddress,nofootinbib,preprintnumbers,notitlepage]{revtex4-2}

\usepackage{amsmath,amssymb,amsthm,amsfonts}
\usepackage{graphicx,tabularx}
\usepackage{color}
\usepackage{multirow}
\usepackage{comment}
\usepackage{enumitem}
\usepackage{subfigure}
\usepackage{orcidlink}
\usepackage{cleveref}
\usepackage{slashed}
\usepackage[normalem]{ulem}
\usepackage{scalerel}
\usepackage{wrapfig}
\usepackage{cancel}
\usepackage{xcolor}
\usepackage{multirow}
\usepackage{setspace}
\usepackage{dcolumn}	
\usepackage{bm}	     	
\usepackage{setspace}
\usepackage{siunitx}
\usepackage{datetime}
\usepackage{fancyvrb}
\usepackage{textgreek}
\usepackage{xspace}
\usepackage{appendix}
\usepackage{adjustbox}
\usepackage{tikz}
\usetikzlibrary{shapes.geometric, arrows}
\usepackage[normalem]{ulem}

\newcommand{\PRE}[1]{{#1}}

\newcommand{\be}{\begin{equation} \begin{aligned}}
\newcommand{\ee}{\end{aligned} \end{equation}}
\newcommand{\beqa}{\begin{eqnarray}}
\newcommand{\eeqa}{\end{eqnarray}}

\def\figureautorefname~#1\null{Fig.\,#1\null}
\def\tableautorefname~#1\null{Tab.\,#1\null}
\def\equationautorefname~#1\null{Eq.\,(#1)\null}

\RequirePackage[normalem]{ulem}

\crefname{section}{Sec.}{Secs.}
\crefname{figure}{Fig.}{Figs.}
\crefname{equation}{Eq.}{Eqs.}
\crefname{table}{Table}{Tables}
\crefname{appendix}{Appendix}{Appendices}

\newcommand{\fasernu}{FASER$\nu$\xspace}

\newcommand{\nsegsum}{$N_{\mathrm{7\text{-}core}}$\xspace}
\newcommand{\nmax}{$N_{\mathrm{max}}$\xspace}
\newcommand{\ntotal}{$N_{\mathrm{total}}$\xspace}
\newcommand{\anglerms}{$\sigma_\theta$}

\newcommand{\mum}{\ensuremath{\,\mu\mathrm{m}}\xspace}

\makeatletter
\renewcommand{\p@subsection}{}
\renewcommand{\p@subsubsection}{}
\makeatother

\usepackage{siunitx}
\usepackage{overpic}
\usepackage{lineno}
\usepackage{booktabs}
\usepackage{multirow}

\begin{document}


\title{{Electromagnetic Shower Reconstruction and Identification in FASER's Emulsion Detector for LHC Forward Neutrino Measurements}
\bigskip \\
{\normalsize FASER Collaboration}
}

\begin{figure*}[h]
\vspace*{-0.6in}
\begin{flushleft}
\includegraphics[width=0.19\textwidth]{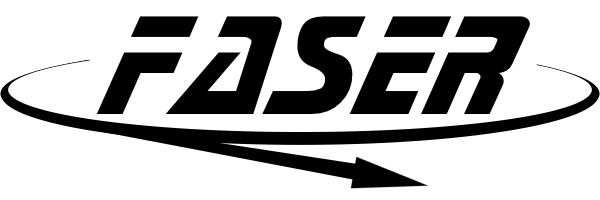}
\end{flushleft}
\end{figure*}

\author{Roshan Mammen Abraham\,\orcidlink{0000-0003-4678-3808}}
\affiliation{Department of Physics and Astronomy, University of California, Irvine, CA 92697-4575, USA}

\author{Xiaocong Ai\,\orcidlink{0000-0003-3856-2415}}
\affiliation{School of Physics, Zhengzhou University, Zhengzhou 450001, China}
  
\author{Saul Alonso Monsalve\,\orcidlink{0000-0002-9678-7121}}
\affiliation{Institute for Particle Physics, ETH Z\"urich, Z\"urich 8093, Switzerland}

\author{John Anders\,\orcidlink{0000-0002-1846-0262}}
\affiliation{University of Liverpool, Liverpool L69 3BX, United Kingdom}

\author{Emma Kate Anderson\,\orcidlink{0000-0002-0161-4560}}
\affiliation{CERN, CH-1211 Geneva 23, Switzerland}

\author{Akitaka Ariga\,\orcidlink{0000-0002-6832-2466}}
\affiliation{Albert Einstein Center for Fundamental Physics, Laboratory for High Energy Physics, University of Bern, Sidlerstrasse 5, CH-3012 Bern, Switzerland}
\affiliation{Department of Physics, Chiba University, 1-33 Yayoi-cho Inage-ku, 263-8522 Chiba, Japan}

\author{Tomoko Ariga\,\orcidlink{0000-0001-9880-3562}}
\affiliation{Kyushu University, 744 Motooka, Nishi-ku, 819-0395 Fukuoka, Japan}

\author{Jeremy Atkinson\,\orcidlink{0009-0003-3287-2196}}
\affiliation{Albert Einstein Center for Fundamental Physics, Laboratory for High Energy Physics, University of Bern, Sidlerstrasse 5, CH-3012 Bern, Switzerland}

\author{Florian~U.~Bernlochner\,\orcidlink{0000-0001-8153-2719}}
\affiliation{Universit\"at Bonn, Regina-Pacis-Weg 3, D-53113 Bonn, Germany}

\author{Jianming Bian\,\orcidlink{0000-0003-3739-5424}}
\affiliation{Department of Physics and Astronomy, University of California, Irvine, CA 92697-4575, USA}

% ...

\begin{abstract}
We present methods for electromagnetic shower reconstruction and identification in the \fasernu emulsion detector using 100\,GeV and 200\,GeV electron test-beam data from the CERN SPS H4 beamline. The reconstruction employs a clustering-based algorithm without energy-dependent tuning to determine shower axes. A multi-level identification chain comprising track pre-selection, a cut-based selection, and a BDT classifier achieves combined background rejection rates of 99.99\% (100\,GeV) and 99.94\% (200\,GeV). The method reaches total reconstruction and identification efficiencies of 58.9\% (100\,GeV) and 70.8\% (200\,GeV) evaluated from simulated samples. Energy reconstruction using the total number of reconstructed segments as the calorimetric estimator yields relative biases of $+0.6\%$ (100\,GeV) and $-0.8\%$ (200\,GeV), with resolutions of 25.4\% and 22.6\%, respectively. Systematic uncertainties on the energy reconstruction are dominated by variations in emulsion film detection efficiency, with totals of $\substack{^{+10.9}_{-8.2}\%}$ at 100\,GeV and $\substack{^{+10.3}_{-6.9}\%}$ at 200\,GeV. The methodology provides a validated framework for electron neutrino identification with the \fasernu detector at the LHC.
\end{abstract}

\maketitle

\begin{center}
\copyright~2026 CERN for the benefit of the FASER Collaboration. Reproduction of this article or parts of it is allowed as specified in the CC-BY-4.0 license.
\end{center}

\clearpage

\section{Introduction}
\label{sec:introduction}

The Forward Search ExpeRiment~\cite{FASER:nu_study,FASER:technical_proposal} (FASER) is designed to search for light, weakly interacting particles and to study high-energy neutrino interactions at the LHC~\cite{LHC}. Located 480\,m downstream of the ATLAS interaction point along the beam collision axis, FASER accesses a kinematic region not covered by the main LHC detectors.

\begin{figure}[htbp]
\centering
\includegraphics[width=0.7\linewidth]{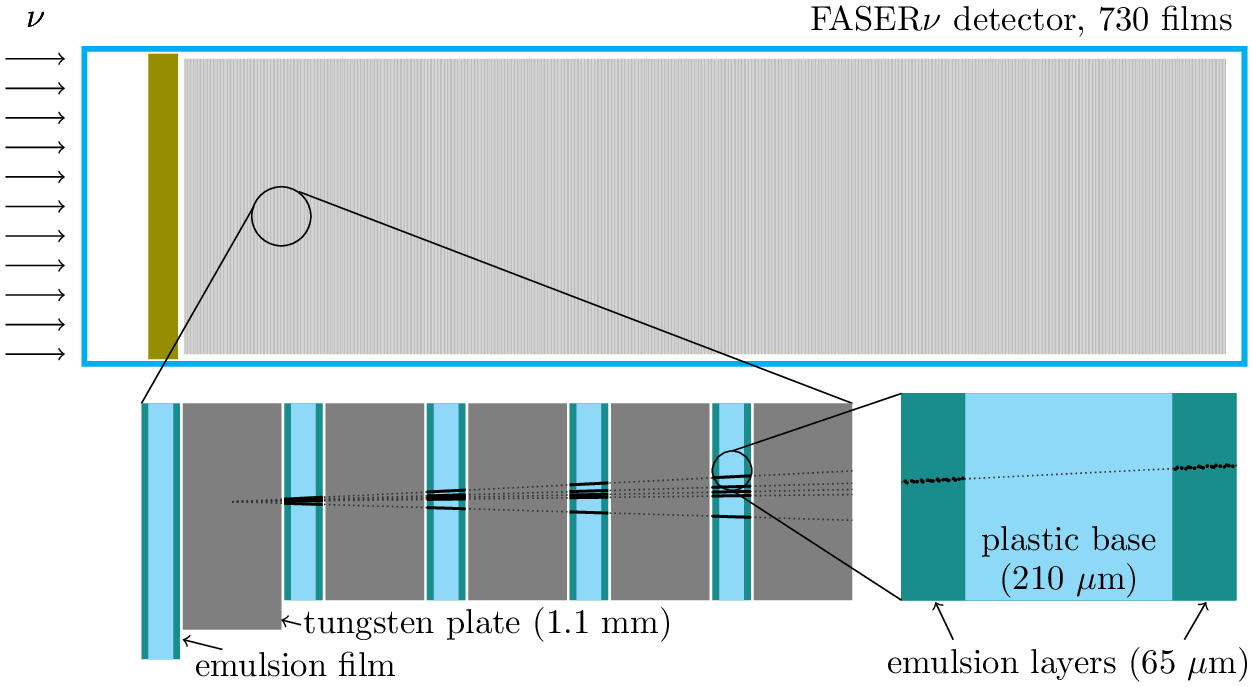}
\caption{Schematic view of the \fasernu detector~\cite{FASER:detector}, consisting of tungsten plates interleaved with OPERA-type emulsion films.}
\label{fig:detector}
\end{figure}

The \fasernu emulsion sub-detector~\cite{FASER:detector,FASER:nu_proposal} serves as both the neutrino interaction target and a high-resolution tracking detector. It has a transverse size of 25 cm $\times$ 30 cm and a length of approximately 100 cm, and consists of 730 tungsten plates (1.1\,mm thick, radiation length $X_0 = 3.5$\,mm) interleaved with double-sided emulsion films~\cite{OPERA:film, ariga2020nuclear}, providing approximately 229 radiation lengths along the beam axis and a total target mass of 1.1\,tonnes. A schematic of the detector is shown in Figure~\ref{fig:detector}.

Each emulsion film comprises two 65\,$\mu$m emulsion layers on each side of a 210\,$\mu$m polystyrene base~\cite{OPERA:film}. Charged particles traversing the emulsion create latent images along their trajectories, which become visible as sequences of silver grains after chemical development. The achieved position resolution is approximately 0.3\,$\mu$m and the angular resolution is approximately 2\,mrad~\cite{HTS}.

The tungsten plates serve a dual purpose: they provide both the neutrino interaction target and the converter material for electromagnetic (EM) shower development. Combined with the fine-grained emulsion tracking, this forms a sampling calorimeter well-suited for reconstructing EM showers from electrons produced in $\nu_e$ charged-current (CC) interactions, as well as photons from $\pi^0$ decays.

EM shower reconstruction and identification are central to the \fasernu physics programme. Electrons from $\nu_e$ CC interactions span energies from approximately 50\,GeV to beyond 2\,TeV~\cite{FASER:nu_rates}, initiating showers through bremsstrahlung and pair production~\cite{PDG} with characteristic high local segment multiplicities. Measuring these electrons in the presence of the intense LHC muon background is essential for determining $\nu_e$ interaction rates and studying neutrino flavour composition. FASER has reported the first neutrino interaction candidates at the LHC~\cite{FASER:nu_candidates}, the first direct observation of collider neutrinos~\cite{FASER:first_nu}, and the first measurements of $\nu_e$ and $\nu_\mu$ cross-sections at TeV energies~\cite{FASER:xsec}. Measuring charged-particle momenta in the emulsion detector using multiple Coulomb scattering has also been demonstrated~\cite{FASER:momentum}. Complementary measurements of forward LHC neutrinos are pursued by the SND@LHC experiment~\cite{SND:first_nu, SND:nunomuon}. The EM shower reconstruction method used in the $\nu_e$ and $\nu_\mu$ cross-section measurement~\cite{FASER:xsec} was optimized for electrons above 200\,GeV and has limitations at lower energies where showers produce fewer segments.

This paper presents improved methods for EM shower reconstruction and identification, validated using test beam data from emulsion modules exposed to 100\,GeV and 200\,GeV electrons at the CERN SPS. The reconstruction uses a density-based clustering algorithm that automatically adapts its parameters from the local segment density, designed to reach energy-independent performance from tens of GeV to multi-TeV. A multi-level identification chain achieves muon background rejection above 99.9\% at both 100\,GeV and 200\,GeV while maintaining high signal efficiency.

The paper is organized as follows. Section~\ref{sec:setup} describes the test-beam configuration and data processing. Section~\ref{sec:reconstruction} presents the shower reconstruction and identification pipeline. Section~\ref{sec:validation} presents performance validation against test-beam data. Energy reconstruction and systematic uncertainties are discussed in Sections~\ref{sec:energy} and~\ref{sec:syst}. Section~\ref{sec:conclusion} summarizes the results.

\section{Experimental Setup and Data Processing}
\label{sec:setup}

\subsection{Test Beam Configuration and Data Quality}

\begin{figure}[htbp]
\centering
\includegraphics[width=0.7\linewidth]{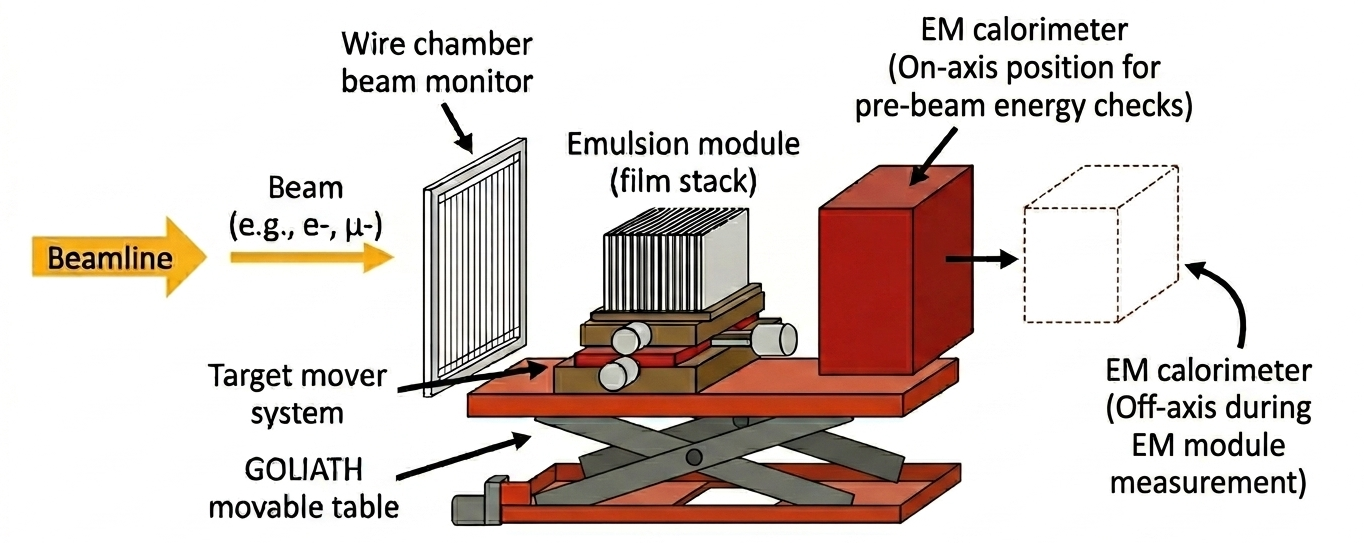}
\caption{Experimental setup at the H4 beamline of the CERN SPS. The emulsion module and electromagnetic calorimeter are mounted on a remotely controlled target mover, which is installed on the GOLIATH movable table downstream of the wire chamber beam monitor. }
\label{fig:setup}
\end{figure}

Emulsion modules were exposed to electron and muon beams at CERN SPS H4 beamline in 2023, providing mono-energetic electron samples at 100\,GeV and 200\,GeV. The experimental setup is shown in Figure~\ref{fig:setup}.

The modules were mounted on a remotely controlled target mover system originally developed for the DsTau experiment~\cite{DsTau}, enabling precise positioning during beam spills to maintain optimal event spacing. The target mover is installed on the GOLIATH movable table. A wire chamber upstream of the target provided beam profile monitoring. Beam intensity was measured by scintillation counters located further upstream in the H4 beamline. An electromagnetic calorimeter provided real-time feedback on beam energy and particle composition; electron purity was measured via calorimeter shower shape discrimination before the emulsion module was moved on-axis for data collection. 

A right-handed coordinate system is adopted throughout this paper: the $Z$-axis points along the beam direction, $X$ is horizontal transverse, and $Y$ is vertical transverse. The achieved beam conditions were:
\begin{itemize}
    \item 100\,GeV (module T1202): 40--50 particles per spill, electron purity $>$80\%;
    \item 200\,GeV (module T1101): 40--60 particles per spill, electron purity $>$90\%.
\end{itemize}
The target exposure density was approximately 8 electrons/cm$^2$. Separate 300\,GeV muon beam exposures provided alignment references, delivering approximately 1500 muons/cm$^2$ to ensure sufficient track statistics for inter-plate alignment across the full module. Beams were oriented perpendicular to the emulsion modules within $\sim2$ mrad in both transverse projections.

After beam exposure, the emulsion films were chemically developed and scanned using the Hyper Track Selector (HTS) system~\cite{HTS}. The trajectory of a charged particle traversing an emulsion layer is reconstructed as a micro-track, a sequence of aligned silver grains. Two corresponding micro-tracks on either side of the plastic base are connected to form a base-track, defined as the straight line linking the grains closest to the two base surfaces. Base-tracks are then connected across consecutive films to reconstruct the full 3-D particle trajectory~\cite{FASER:reco}. Throughout this analysis, a reconstructed track is required to consist of at least three base-tracks; shorter fragments are reconstruction artifacts and are not counted. Before analysis, a track reconnection algorithm is applied to merge broken tracks and improve muon rejection through track-level pre-selection (Section~\ref{sec:presel}).

The measured data quality metrics are summarized in Table~\ref{tab:dataquality}; the underlying film-by-film measurements are presented in Appendix~\ref{app:dataquality}. The 200\,GeV module has systematically higher single-film efficiency ($\sim$95\%) than the 100\,GeV module ($\sim$85\%), attributed to differences in emulsion quality or processing conditions between modules. These efficiency differences are incorporated in Monte Carlo (MC) simulation and propagated as systematic uncertainties (Section~\ref{sec:syst}). The position and angular resolutions are broader than those reported in Refs.~\cite{HTS, FASER:reco} because the fine local alignment stage used in standard \fasernu analyses was not needed in this EM shower test-beam study, and the quoted values thus include the contribution from film-to-film alignment uncertainties.

\begin{table}[htbp]
\centering
\caption{Measured data quality metrics for both test beam modules, evaluated using the methods described in Ref.~\cite{FASER:reco}.}
\label{tab:dataquality}
\begin{tabular}{lcc}
\hline
Parameter & 100\,GeV (T1202) & 200\,GeV (T1101) \\
\hline
Position resolution & 0.5--3.0\,$\mu$m & 0.5--1.5\,$\mu$m \\
Angular resolution ($\sigma_{\tan\theta_x}$) & 2.29\,mrad & 2.45\,mrad \\
Angular resolution ($\sigma_{\tan\theta_y}$) & 1.51\,mrad & 1.44\,mrad \\
Single-film efficiency (peak) & $\sim$85\% & $\sim$95\% \\
\hline
\end{tabular}
\end{table}

\subsection{Monte Carlo Simulation}
\label{sec:mc}

Monte Carlo simulation is used to evaluate signal efficiency, optimize selection criteria, and calibrate the energy estimator. Electron showers are generated using GEANT4~\cite{Geant4} with a detailed geometry model of the emulsion modules, including the tungsten plates, emulsion films, and all passive structural material. The simulation models electromagnetic physics processes including bremsstrahlung, pair production, and ionisation energy loss, as well as hadronic and photonuclear processes relevant to secondary particle production in dense material.
 
Simulated events are processed through the same emulsion track reconstruction chain as data: base-track formation, film-to-film alignment, multi-plate track linking, and track reconnection, with the same single-film efficiency as measured in data: 85\% for the 100\,GeV module and 95\% for the 200\,GeV module. For the 200\,GeV module, the close-pair inefficiency correction is additionally applied; this is discussed in detail in Section~\ref{sec:energy}.

The signal MC samples do not include muon backgrounds, which are instead taken directly from test beam data. This approach is described further in Section~\ref{sec:mediumid}.
 
\section{Shower Reconstruction and Identification}
\label{sec:reconstruction}

\begin{figure}[htbp]
\centering
\includegraphics[width=0.9\textwidth]{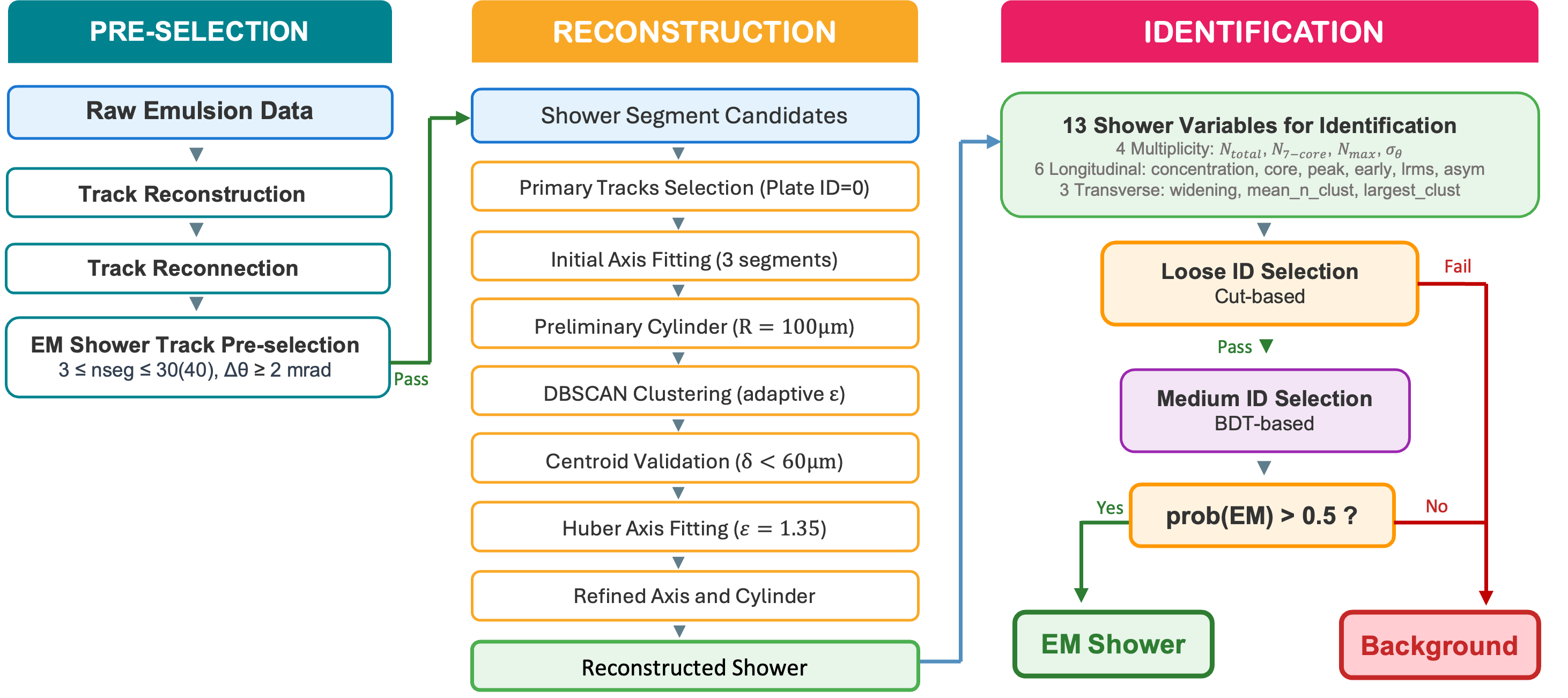}
\caption{Overview of the EM shower reconstruction and identification pipeline, showing the three main stages: pre-selection, reconstruction, and identification.}
\label{fig:flowchart}
\end{figure}

Figure~\ref{fig:flowchart} provides an overview of the complete analysis chain. The pipeline proceeds in three stages: pre-selection applies track-level quality cuts to reject obvious muon backgrounds; shower reconstruction uses adaptive DBSCAN (Density-Based Spatial Clustering of Applications with Noise) clustering~\cite{DBSCAN} and Huber robust axis fitting to collect shower segments within a cylindrical volume; identification applies cut-based Loose ID followed by BDT-based Medium ID to distinguish genuine electrons from residual backgrounds.

\subsection{Preliminary Selection on Tracks}
\label{sec:presel}

Muon tracks traversing the full detector depth are characterized by large segment counts and small scattering angles, whereas EM shower primary tracks are short (initiating cascades within a few radiation lengths) and exhibit larger apparent scattering due to early pair production and bremsstrahlung.

The scattering angle $\theta_s$ is computed for each track. For tracks with four or more segments, independent linear fits are performed on the first three and last three segments, giving slope estimates $T_{X,1}$, $T_{Y,1}$ and $T_{X,2}$, $T_{Y,2}$; the scattering angle is:
\begin{equation}
\theta_s = \sqrt{(T_{X,2} - T_{X,1})^2 + (T_{Y,2} - T_{Y,1})^2}\,.
\end{equation}
For tracks with exactly three segments, $\theta_s$ is computed from the angular difference between the first and last segments.

The pre-selection criteria are:
\begin{itemize}
    \item $3 \leq n_{\mathrm{seg}} \leq 30$ for the 100\,GeV module (T1202);
    \item $3 \leq n_{\mathrm{seg}} \leq 40$ for the 200\,GeV module (T1101);
    \item $\theta_s \geq 2$\,mrad
\end{itemize}
where $n_{\mathrm{seg}}$ denotes the number of segments belonging to a single 3-D track. The lower bound on the segment count removes fragmented segments and implicitly imposes a track momentum threshold of approximately 0.8 GeV, while the upper bound removes muon tracks traversing most of the detector. The different upper bounds at the two energies reflect a module-dependent tracking quality effect and are determined by MC simulation samples. The scattering angle cut rejects through-going muons. After track reconnection and pre-selection, approximately 29\% of tracks originating from the first plate survive at both energies in test-beam data.

\begin{figure}[htbp]
\centering
\includegraphics[width=0.8\textwidth]{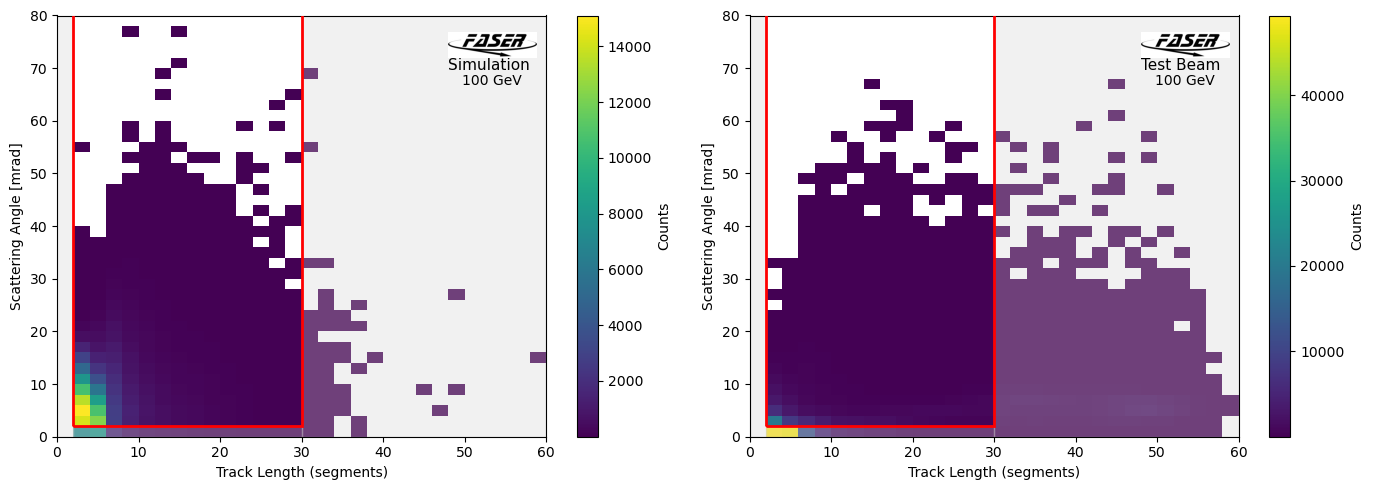}
\includegraphics[width=0.8\textwidth]{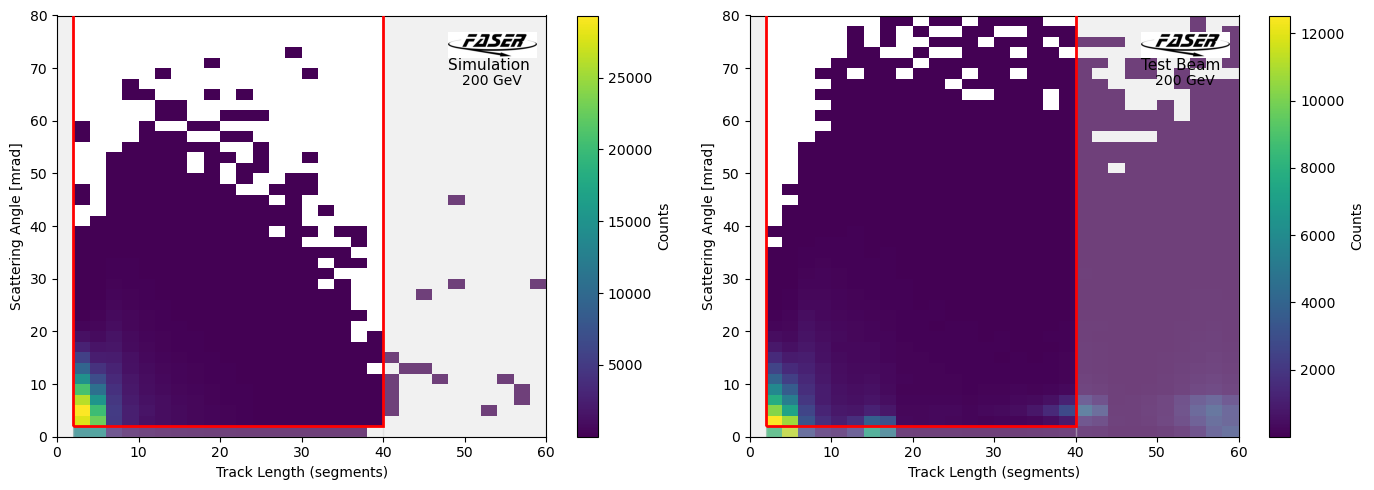}
\caption{Joint distributions of $\theta_s$ and $n_\mathrm{seg}$ for electron MC (left) and test-beam data (right) in the 100\,GeV (top) and 200\,GeV (bottom) modules, after track reconnection. The red rectangle indicates the pre-selection region.}
\label{fig:preselection}
\end{figure}

Figure~\ref{fig:preselection} shows the joint distributions of $\theta_s$ and $n_\mathrm{seg}$ for electron MC and test-beam data in both modules: muon tracks populate the region of large $n_\mathrm{seg}$ and small $\theta_s$, well separated from the short, large-scattering tracks characteristic of EM showers.

\subsection{Electromagnetic Shower Reconstruction}

\label{sec:initaxis}

For the test-beam configuration, where electrons enter the detector at the front face, reconstruction begins with tracks originating from the first emulsion plate. MC studies show that 79.4\% (100\,GeV) and 89.8\% (200\,GeV) of primary electrons have their first reconstructed segment on plate 1; the remaining fraction typically appear within the first five plates due to insufficient ionization or local detector and reconstruction inefficiency.

The initial shower axis is determined by fitting the first three segments of each primary track independently in the Z-X and Z-Y projection planes using linear regression. Using three plates provides optimal angular resolution ($\sigma \approx 0.5$--$0.8$\,mrad) while minimizing deflections from early shower development and multiple scattering; using fewer plates gives insufficient constraints, while more plates increasingly biases the fit through shower scattering.

A preliminary cylindrical search volume is then defined around the initial axis, with radius 100\mum and length 8\,cm (approximately 18 radiation lengths, spanning 56 plates). Segments are included in the preliminary shower candidate if they belong to a track with at least 3 segments ($N_{\mathrm{seg}} \geq 3$), have a maximum perpendicular distance from the axis $\leq 100$\mum, a maximum angular deviation between the segment direction and the initial axis direction $\leq 10$\,mrad, and a minimum distance of closest approach between the segment trajectory and the initial axis $\leq 50$\mum. These criteria provide the input segment sample for DBSCAN clustering.

\label{sec:dbscan}

Within the preliminary cylindrical volume, DBSCAN is applied independently to segments on each plate using their 2D spatial coordinates (X, Y). It groups points in high-density regions while classifying isolated points as noise, making it well-suited for separating the dense EM shower core from background tracks and peripheral shower particles. Unlike $k$-means~\cite{kmeans}, DBSCAN does not require the number of clusters to be specified in advance and can identify clusters of arbitrary shape. The algorithm requires two input parameters: a neighbourhood radius $\varepsilon$, within which points are considered neighbours, and a minimum cluster size $N_{\rm min}$, the minimum number of neighbouring points required to form a cluster.

The key feature of our implementation is adaptive parameter selection based on local segment density, without requiring prior knowledge of the shower energy. The neighbourhood radius $\varepsilon$ is estimated from the median nearest-neighbour distance among segments on each plate:
\begin{equation}
\varepsilon = 2.5 \times \mathrm{median}(d_{\mathrm{NN}})
\end{equation}
where $\mathrm{median}(d_{\mathrm{NN}})$ is the median nearest-neighbour distance over all segments on the plate. This value is bounded within 10--80\mum: the lower bound reflects the close-pair reconstruction limit ($\sim$10\mum) where overlapping grain patterns cause reconstruction inefficiency, and the upper bound corresponds to typical transverse shower dimensions.

The minimum cluster size $N_{\mathrm{min}}$ scales with the total segment count $N_{\mathrm{seg}}$ on the plate:
\begin{equation}
N_{\mathrm{min}} = \max(2,\, \lfloor N_{\mathrm{seg}}/4 \rfloor)
\end{equation}
For sparse plates early in shower development, $N_{\mathrm{min}} = 2$ permits cluster formation with minimal requirements. For denser plates near the shower maximum, requiring approximately 25\% of segments ensures the shower core forms a single dense cluster while rejecting isolated background segments.

\begin{figure}[htbp]
\centering
\includegraphics[width=0.5\linewidth]{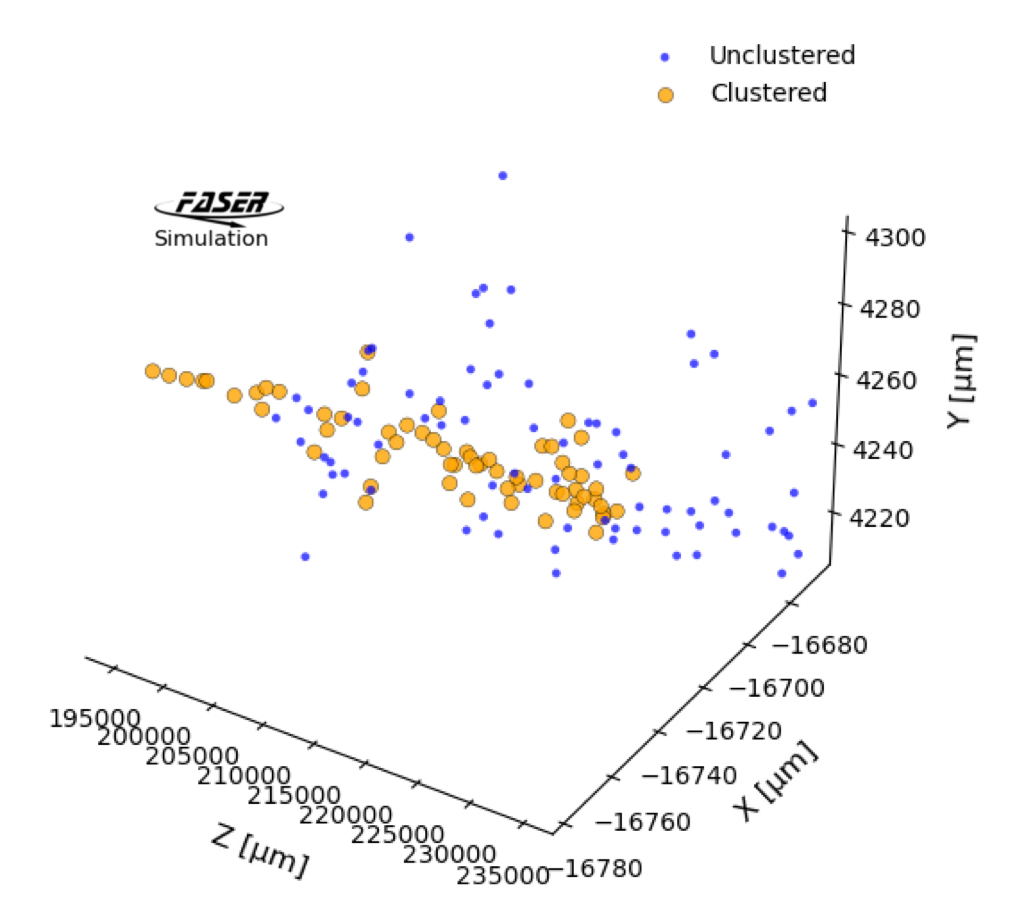}
\caption{Example of DBSCAN clustering on a 200\,GeV electron MC shower. The identified shower core cluster (yellow) is separated from peripheral shower particle segments (blue). On the first plates, the segments used for the initial axis fit are assigned to the core cluster directly and therefore appear as clustered.}
\label{fig:dbscan}
\end{figure}

When multiple clusters are identified on a plate, the cluster nearest to the expected shower position extrapolated from the current axis estimate is selected. For each selected cluster, the spatial centroid is computed:
\begin{equation}
(\bar{X}, \bar{Y}, \bar{Z}) = \frac{1}{N_{\mathrm{cluster}}} \sum_{i=1}^{N_{\mathrm{cluster}}} (X_i, Y_i, Z_i)
\end{equation}
Figure~\ref{fig:dbscan} shows an example of DBSCAN clustering on a shower from the 200\,GeV electron MC sample.

\label{sec:axisfit}

To suppress contamination from peripheral shower particles that may create spurious centroids far from the true shower axis, a sequential validation procedure is applied before the final axis fit. The first-plate centroid is always accepted as the shower entry point. For each subsequent plate $i > 1$, a provisional axis is fit using all previously validated centroids via linear regression, and the deviation of the current centroid from this provisional axis is computed:
\begin{equation}
\delta_i = \sqrt{(X_i - X_{\mathrm{fit}}(Z_i))^2 + (Y_i - Y_{\mathrm{fit}}(Z_i))^2}\,.
\end{equation}
A centroid is accepted if $\delta_i < 60$\mum; otherwise it is rejected and excluded from subsequent fitting. Additionally, DBSCAN clustering is applied only up to the $\text{30}^{\text{th}}$ plate, focusing on the main shower development region. The self-correcting nature of this procedure ensures that early validated plates establish a reliable axis direction, and the constraint becomes increasingly effective as more validated centroids accumulate.

\begin{figure}[htbp]
\centering
\includegraphics[width=0.8\linewidth]{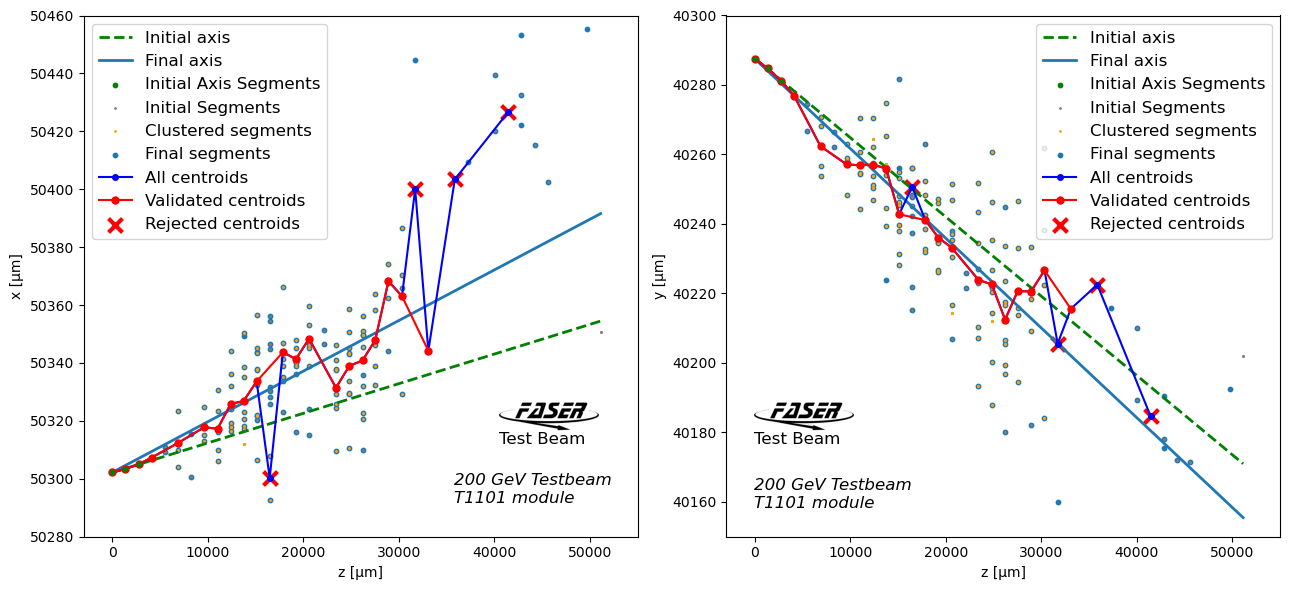}
\caption{Complete reconstruction chain for a representative 200\,GeV test-beam shower candidate in Z-X (left) and Z-Y (right) projections. The initial axis (dashed green) is refined through DBSCAN centroid fitting and Huber regression to yield the final axis (solid blue). Red circles: validated centroids; red crosses: rejected centroids. Gray points: all segments in the preliminary cylinder. Blue points: segments in the final reconstructed shower.}
\label{fig:finalshower}
\end{figure}

The validated centroids are then fit using Huber regression~\cite{Huber} with an entry-point constraint. The Huber loss function is applied to the residual $r$, defined as the deviation of a centroid from the fitted axis in a given projection plane:
\begin{equation}
L_\delta(r) = \begin{cases}
\frac{1}{2}r^2 & \text{for } |r| \leq \delta \\
\delta\!\left(|r| - \tfrac{1}{2}\delta\right) & \text{for } |r| > \delta
\end{cases}
\end{equation}
The loss behaves quadratically for small residuals and linearly for large residuals, reducing the influence of occasional spurious centroids compared to ordinary least squares. The threshold $\delta = \varepsilon \cdot \hat{\sigma}$ is determined adaptively from the estimated residual scale $\hat{\sigma}$ of each fit, using $\varepsilon = 1.35$ as the scikit-learn~\cite{scikit-learn} default. The entry-point constraint is implemented by translating the coordinate origin to the first centroid before fitting, anchoring the axis at the shower initiation point. The fit is performed independently in the Z-X and Z-Y projection planes:
\begin{equation}
X = a_X \cdot Z + b_X, \quad Y = a_Y \cdot Z + b_Y
\end{equation}
where the slopes $a_X = \tan\theta_X$ and $a_Y = \tan\theta_Y$ define the refined shower direction. Using this refined axis, a final cylindrical volume with the same geometric criteria as the preliminary selection (radius 100\mum, angular deviation $< 10$\,mrad, minimum approach distance $< 50$\mum) is defined. All segments within this cylinder constitute the reconstructed shower and are used for subsequent profile analysis and energy reconstruction. The complete reconstruction chain for a test-beam event is illustrated in Figure~\ref{fig:finalshower}.

\begin{figure}[htbp]
\centering
\includegraphics[width=0.8\linewidth]{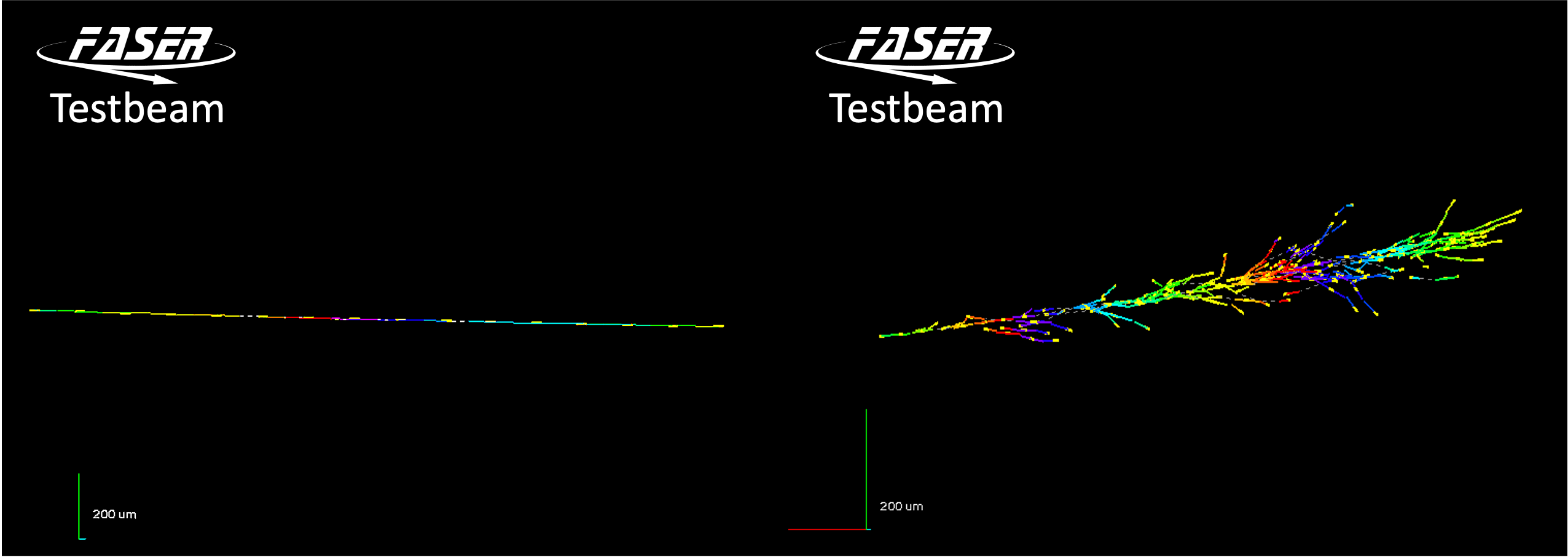}
\caption{Representative reconstructed events from the 200\,GeV test beam module: a muon track (left) and an electromagnetic shower (right).}
\label{fig:eventdisplay}
\end{figure}

Figure~\ref{fig:eventdisplay} shows representative reconstructed events from the test beam data. The two particle types are morphologically distinct: a muon appears as a single straight track with typically one segment per plate, whereas an EM shower initiates a dense cascade with rapidly increasing segment multiplicity across successive plates, significant transverse spread around the shower axis, and gradual attenuation downstream. These topological differences are used in two contexts in this analysis through visual inspection of emulsion event displays. For the BDT background training sample (Section~\ref{sec:mediumid}), muon candidates are selected by visually confirming the absence of shower-like topology. The same visual confirmation is applied when measuring muon background rejection rates in test beam data (Section~\ref{sec:validation}), where tracks surviving the identification chain are inspected to verify whether they are genuine muon backgrounds or misidentified EM showers. Both selections are performed by trained scanners familiar with emulsion event topologies.
 
\subsection{Electromagnetic Shower Identification}

With the shower segments collected and the two topologies established, the next step is to distinguish genuine EM showers from the dominant muon background. In the test-beam environment, muons constitute more than 99\% of all reconstructed tracks, requiring rejection factors exceeding $10^3$ to achieve acceptable signal purity. We employ a multi-level identification strategy that progressively refines the sample through increasingly discriminating selections, ensuring the BDT classifier operates on a pre-cleaned sample where obvious backgrounds have already been removed.

\label{sec:looseid}

After shower reconstruction, a cut-based selection (Loose ID) removes the bulk of remaining muon tracks that mimic shower topologies through accidental segment accumulation in the cylindrical volume. Table~\ref{tab:looseid} lists the selection criteria and the definitions of all variables used.
 
\begin{table}[htbp]
\centering
\caption{Loose ID selection criteria. $N_{\mathrm{max}}$: maximum segment count on any single plate; $N_{\mathrm{7\text{-}core}}$: segments in the seven plates centred on the shower maximum; $N_{\mathrm{total}}$: total reconstructed segments across all plates; $\sigma_\theta$: angular spread RMS of shower segments (Eq.~\ref{eq:sigtheta}). Plates are numbered sequentially along the beam direction, with plate 1 the most upstream and plate 60 the most downstream.}
\label{tab:looseid}
\renewcommand{\arraystretch}{1.2}
{\setlength{\tabcolsep}{8pt}
\begin{tabular}{llc}
\hline
Variable & Definition & Requirement \\
\hline
\nmax & Maximum segment count on any single plate & $\geq 4$ \\
Shower max plate & Plate ID of \nmax & $\geq 3$ \\
Shower start plate & First plate with more than one shower segment & $\leq 20$ \\
$N_{\mathrm{7\text{-}core}}$ & Segments in seven plates centred on Shower max plate & $\geq 10$ \\
\ntotal & Total reconstructed segments across all plates & $\geq 20$ \\
$\sigma_\theta$ & Angular spread RMS of shower segments & $\geq 4$\,mrad \\
\hline
\end{tabular}}
\end{table}
 
where $\sigma_\theta$ is defined as:
\begin{equation}
\sigma_\theta = \sqrt{\frac{1}{N}\sum_{i=1}^{N}\left[(\Delta\theta_{x,i})^2 + (\Delta\theta_{y,i})^2\right]}
\label{eq:sigtheta}
\end{equation}
the RMS of the angular deviations of all $N$ shower segments from their mean direction. 

The Loose ID selection achieves signal efficiencies of 91.7\% (100\,GeV) and 99.7\% (200\,GeV) on MC, and muon background rejection rates of 98.7\% (100\,GeV) and 99.9\% (200\,GeV) in test-beam data. At 200\,GeV this level of rejection is already sufficient for many analyses; at 100\,GeV the more challenging signal-to-background ratio motivates the additional Medium ID step.

\label{sec:mediumid}

The BDT-based Medium ID provides additional discrimination using ten shower shape variables. Multiplicity variables such as \nsegsum or $N_{\mathrm{max}}$ vary strongly with energy and would require separate tuning at different energies. We instead use the shower shape variables, which do not scale with the overall segment multiplicity and therefore vary only weakly with energy. The ten input features are mainly constructed from the quantities used in LooseID: total segments \ntotal, shower core segments \nsegsum, shower max \nmax, angular spread RMS \anglerms, plus the travel distance $L$ (number of plates over which shower segments are detected), and are grouped into six longitudinal variables, three transverse and clustering variables, and one angular variable, as described below.

\textbf{Six longitudinal variables:}
\begin{enumerate}
    \item \textit{Concentration Ratio} ($N_{\mathrm{7\text{-}core}}/N_{\mathrm{total}}$): fraction of segments within the seven-plate core region centered on shower maximum.
    \item \textit{Core Fraction} ($N_{\mathrm{7\text{-}core}}/L$): \nsegsum normalized by the number of plates traversed.
    \item \textit{Peak Fraction} ($N_{\mathrm{max}}/N_{\mathrm{total}}$): fraction of total segments on the single most active plate.
    \item \textit{Early Fraction} ($N_{\mathrm{first\,half}}/N_{\mathrm{total}}$): fraction of segments in the first half of plates traversed.
    \item \textit{Longitudinal RMS}: segment-count-weighted RMS of plate positions,
    \begin{equation}
    \bar{p} = \frac{\sum_j n_j p_j}{\sum_j n_j}, \quad\sigma_L = \sqrt{\frac{\sum_j n_j (p_j - \bar{p})^2}{\sum_j n_j}}
    \end{equation}
    where $n_j$ is the number of segments on plate $j$ and $p_j$ is the plate number.
    \item \textit{Asymmetry}: imbalance of segments before versus after the shower maximum plate,
    \begin{equation}
    A = \frac{|N_{\mathrm{before}} - N_{\mathrm{after}}|}{N_{\mathrm{before}} + N_{\mathrm{after}}}
    \end{equation}
\end{enumerate}

\textbf{Three transverse and clustering variables:}
\begin{enumerate}
    \setcounter{enumi}{6}
    \item \textit{Widening Ratio}: ratio of the mean transverse distance from the reconstructed axis in the shower maximum region ($\pm$1 plate) to that in the early region (plates before shower maximum $-3$), with 1\mum added to the denominator for regularization.
    \begin{equation}
    W = \frac{\langle d_{\perp} \rangle_{\mathrm{max}}}{\langle d_{\perp} \rangle_{\mathrm{early}} + 1\,\mu\mathrm{m}}
    \end{equation}
    where $\langle d_{\perp} \rangle$ denotes the mean perpendicular distance of shower segments from the reconstructed axis in the respective region.
    \item \textit{Mean Cluster Number}: average number of DBSCAN clusters identified per plate, measuring how fragmented the shower appears across plates.
    \item \textit{Mean Largest-Cluster Fraction}: fraction of segments belonging to the largest DBSCAN cluster on each plate, averaged over all plates.
\end{enumerate}

\textbf{One angular variable:}
\begin{enumerate}
    \setcounter{enumi}{9}
    \item \textit{Angular Spread RMS} ($\sigma_\theta$): RMS of the angular deviations $\sqrt{(\Delta\theta_x)^2 + (\Delta\theta_y)^2}$ of all shower segments from their mean direction, measuring the overall angular divergence of tracks within the shower.
\end{enumerate}

The angular spread RMS is the only variable that participates in both Loose ID and Medium ID. It provides a clear separation between muon-like tracks (typically centered around 2\,mrad) and shower-like candidates (above $\sim$4\,mrad), and contributes approximately 30--40\% (measured by feature importance gain) of the BDT discrimination power. Unlike the absolute variables, $\sigma_\theta$ remains stable across energies from 20 to 300\,GeV, making it particularly robust across the \fasernu energy range.

\begin{figure}[htbp]
\centering
\includegraphics[width=\textwidth]{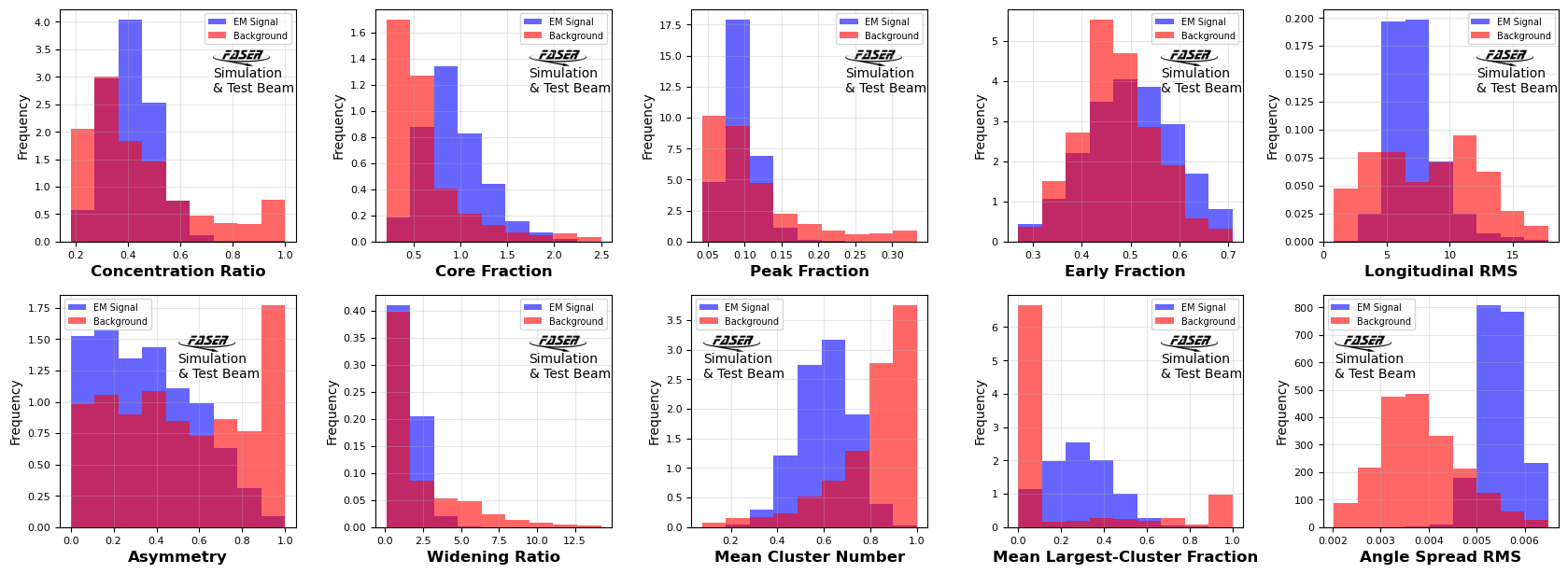}
\caption{Distributions of the ten BDT input variables for EM shower signal (blue, from 100\,GeV electron MC) and muon background (red, from test-beam data identified by visual inspection). The six longitudinal variables provide strong individual separation; the three transverse and clustering variables contribute complementary information for the most ambiguous backgrounds; the angular spread RMS provides robust separation across energies.}
\label{fig:features}
\end{figure}

Figure~\ref{fig:features} shows the distributions of all ten variables for signal (from 100\,GeV electron MC) and background (from test-beam data identified by visual inspection). The corresponding distributions at 200\,GeV, which are not used for training, and the dependence of all ten variables on energy are shown in
Appendix~\ref{app:idvars}.

A Boosted Decision Tree classifier is trained using XGBoost~\cite{XGBoost} with 200 trees, maximum depth 4, learning rate 0.05, and subsample fraction 0.7, optimized through 5-fold cross-validation. The training sample consists of 2517 electrons from MC simulation processed through the full reconstruction chain as signal, and 3517 muon candidates from test beam data identified by visual inspection as background. The classifier achieves an area under the ROC curve (AUC) of 0.997, where 1.0 corresponds to perfect separation.

\begin{figure}[htbp]
\centering
\includegraphics[width=\textwidth]{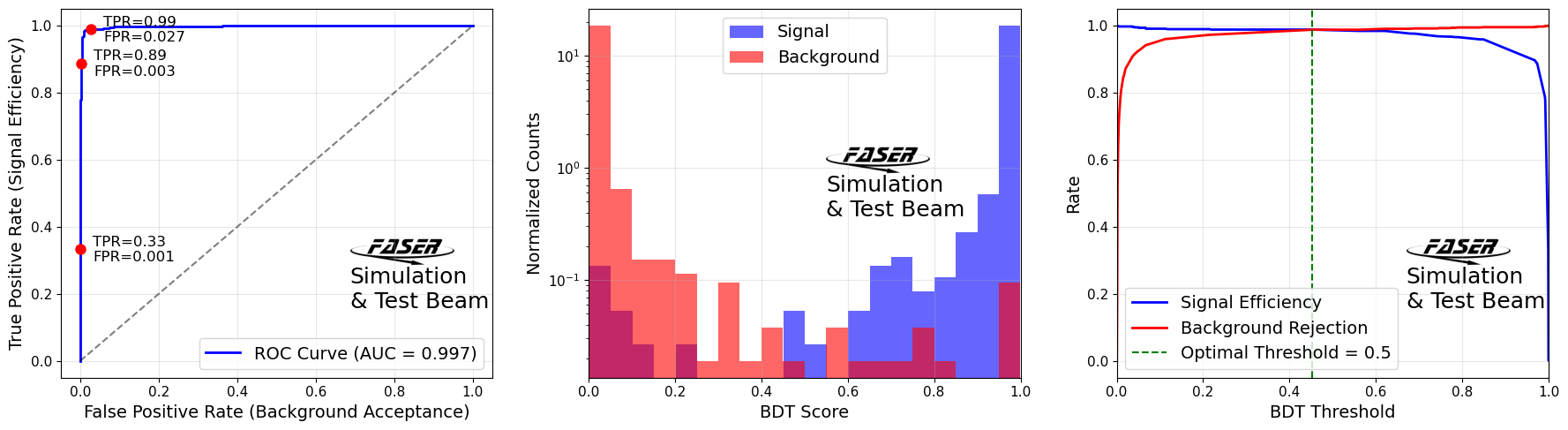}
\caption{BDT classifier performance: ROC curve with AUC = 0.997 (left), BDT score distributions for signal (blue, from 100\,GeV electron MC) and background (red, from test-beam data identified by visual inspection) (center), and signal efficiency and background rejection as a function of BDT threshold (right). The chosen threshold BDT $> 0.5$ is indicated.}
\label{fig:bdt}
\end{figure}

Figure~\ref{fig:bdt} shows the classifier performance. A classification threshold of BDT $> 0.5$ is adopted, chosen to maximize background rejection while maintaining high signal efficiency.

\section{Performance Validation}
\label{sec:validation}

Tables~\ref{tab:id_performance} summarizes the step-by-step efficiency of the full reconstruction and identification chain, where each entry is the fraction of events passing that step relative to all events entering it. Table~\ref{tab:bkg_rejection} gives the corresponding background rejection at each identification stage, which is evaluated on the muon background sample identified by visual inspection in test-beam data. Combined background rejection exceeds 99.9\% at both energies. At 100\,GeV, the expected muon density of $\sim$1500\,cm$^{-2}$ compared to $\sim$8\,cm$^{-2}$ electrons means that Loose ID alone (98.66\% rejection) leaves a residual background that significantly outnumbers the signal; the Medium ID provides the additional discrimination needed to reach 99.99\% total rejection. At 200\,GeV, Loose ID already achieves 99.87\% rejection, leaving very few muon tracks for the Medium ID to act on; consequently the Medium ID rejection at 200\,GeV (53.85\%) is based on small statistics and carries larger statistical uncertainty, though the combined total rejection of 99.94\% is well determined. Depending on the specific background conditions of an analysis, users may choose to apply either the Loose ID or the full Medium ID selection.

\begin{table}[htbp]
\centering
\caption{Selection efficiency at each stage of the reconstruction and identification chain, defined as the fraction of events passing each selection step. Signal efficiencies are from MC simulation; all-events efficiencies are from test-beam data.}
\label{tab:id_performance}
\begin{tabular}{lcccc}
\hline
 & \multicolumn{2}{c}{100\,GeV} & \multicolumn{2}{c}{200\,GeV} \\
Selection step & Signal (MC) & All events (data) & Signal (MC) & All events (data) \\
\hline
Event starting at 1st plate          & 79.4\% & ---    & 89.8\% & ---    \\
Primary track passing pre-selection  & 80.5\% & 28.8\% & 83.2\% & 28.9\% \\
Reconstructed events                 & 99.9\% & 54.5\% & 96.5\% & 55.5\% \\
Events passing Loose ID               & 95.3\% & 2.66\% & 98.4\% & 2.32\% \\
Events passing Medium ID             & 96.8\% & 43.1\% & 99.9\% & 96.4\% \\
\hline
\textbf{Total Efficiency} & \textbf{58.9\%} & \textbf{0.18\%} & \textbf{70.8\%} & \textbf{0.36\%} \\
\hline
\end{tabular}
\end{table}

\begin{table}[htbp]
\centering
\caption{Background rejection rate at each stage of the identification chain, evaluated on the muon background sample identified by visual inspection in test-beam data.}
\label{tab:bkg_rejection}
\begin{tabular}{lcc}
\hline
Selection step & 100\,GeV & 200\,GeV \\
\hline
Rejected by Loose ID   & 98.66\% & 99.87\% \\
Rejected by Medium ID & 99.26\% & 53.85\% \\
\hline
\textbf{Total background rejection} & \textbf{99.99\%} & \textbf{99.94\%} \\
\hline
\end{tabular}
\end{table}

We then validate the reconstruction and identification chain by comparing key shower observables across three samples: MC with true axis (using the generator-level primary electron direction for cylinder selection), MC with reconstructed axis (using the full reconstruction algorithm), and test-beam data after the full identification chain with BDT\,$>$\,0.5. 

\begin{figure}[htbp]
\centering
\includegraphics[width=\textwidth]{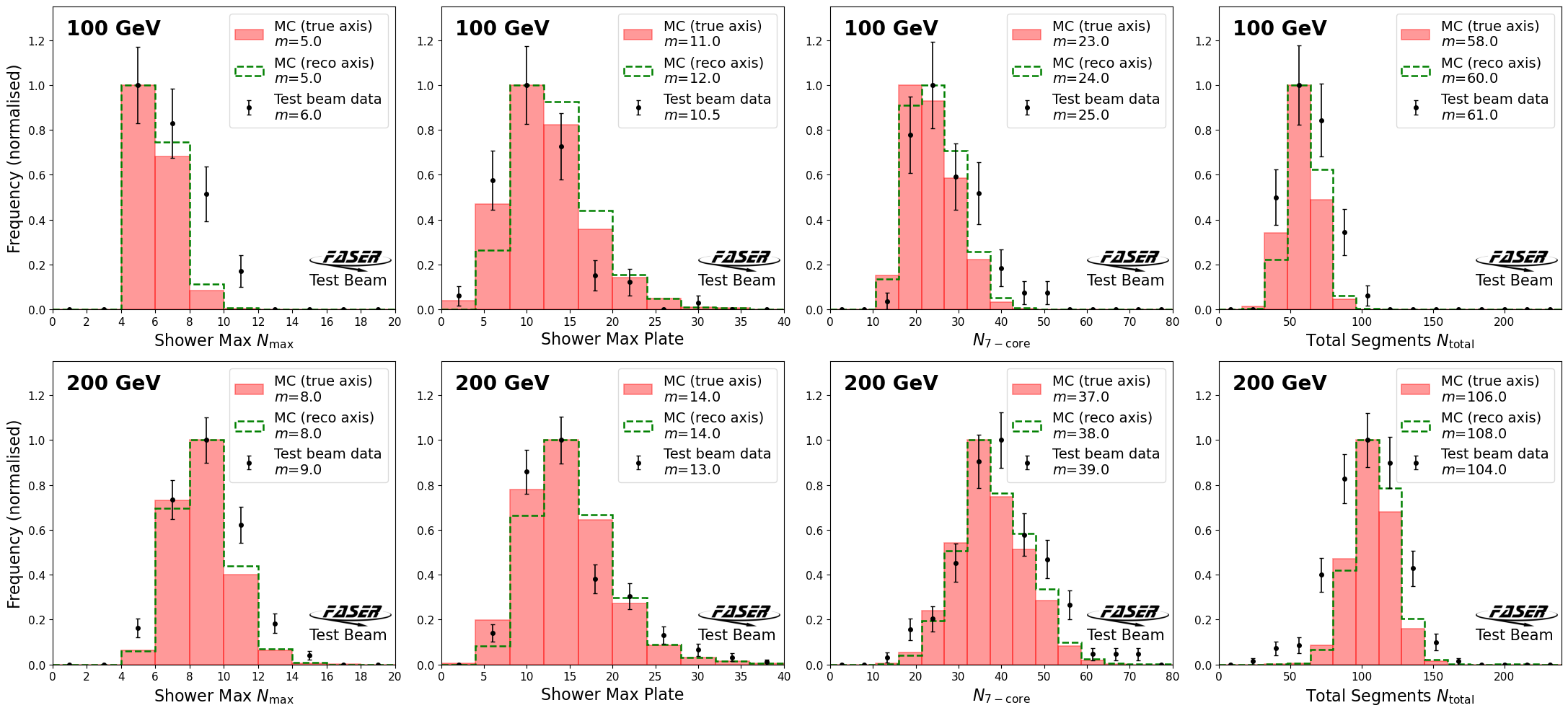}
\caption{Distributions of shower observables at 100\,GeV (top row) and 200\,GeV (bottom row): Shower Max (left), Shower Max Plate (center-left), \nsegsum (center-right), and Total Segments (right). Red filled: MC with true axis; green outline: MC with reconstructed axis; black points with error bars: test-beam data after full identification chain. MC uses single-film efficiencies of 85\% (100\,GeV) and 95\% (200\,GeV).}
\label{fig:shower_observables}
\end{figure}

Comparing MC with true and reconstructed axes quantifies reconstruction-induced bias; comparing MC with reconstructed axis to test-beam data validates the accuracy of the detector simulation. Figure~\ref{fig:shower_observables} shows the four primary shower observables at both energies.

MC distributions with true and reconstructed axes agree closely at both energies, confirming that DBSCAN-based reconstruction recovers shower properties with minimal bias. Shower Max peaks at $m = 5$ (100\,GeV) and $m = 8$ (200\,GeV) in MC, consistent with test-beam data at $m = 6$ and $m = 9$, respectively. Shower Max Plate shows a consistent forward shift of approximately one plate in test-beam data relative to MC at both energies (median 10.5 vs.\ 12 at 100\,GeV; 13 vs.\ 14 at 200\,GeV). This offset is attributed to upstream material not fully modelled in the simulation geometry; it is consistent between both energies and does not affect the shower shape variables used for identification or energy reconstruction. \nsegsum agrees well between MC and data at both energies in both peak position and distribution shape (median 25 vs.\ 24 at 100\,GeV; 39 vs.\ 38 at 200\,GeV), validating that the seven-plate shower core is reliably captured by the reconstruction. For Total Segments, the medians are consistent between data and MC at 100\,GeV (61 vs.\ 60). At 200\,GeV, test-beam data (median 104) is approximately 4\% lower than MC (median 108), with the deficit concentrated in the shower head and tail while the \nsegsum core region agrees well. This discrepancy is attributed primarily to differences in track pre-selection efficiency between data and MC, and is accounted for in the systematic uncertainties (Section~\ref{sec:syst}).

\begin{figure}[htbp]
\centering
\includegraphics[width=0.8\textwidth]{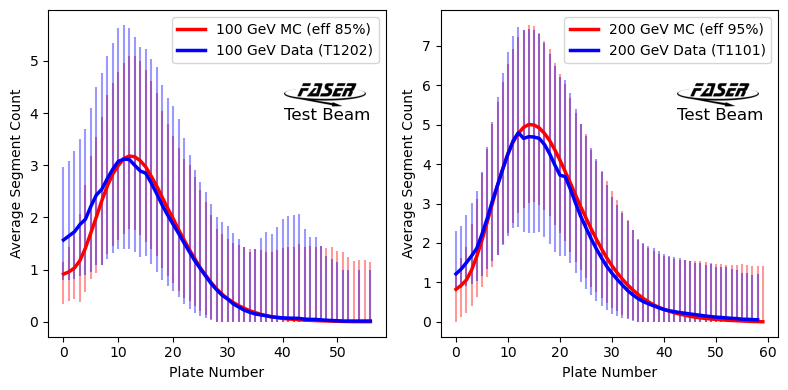}
\caption{Averaged longitudinal shower profiles from MC (red) and test beam data (blue) for 100\,GeV (left) and 200\,GeV (right). Each point shows the mean segment count per plate averaged over all reconstructed showers; error bars show the asymmetric RMS spread of segment counts across showers, computed separately for deviations above and below the mean. MC single-film efficiency is 85\% at 100\,GeV and 95\% at 200\,GeV.}
\label{fig:profile}
\end{figure}

The averaged longitudinal shower profiles are shown in Figure~\ref{fig:profile}, showing good agreement at both energies, which confirms that the shower development physics is well modelled. The residual structure in the data profiles reflects the film-to-film variation of the single-film efficiency, which has a spread of
$\sigma = 6.7\%$ at 100\,GeV and $3.9\%$ at 200\,GeV (Appendix~\ref{app:dataquality}), whereas the simulation assumes a single
uniform efficiency per module. Unlike the integrated segment count used for
energy reconstruction, the per-plate profile does not average over films, so
this variation propagates directly into the plate-to-plate structure; it is
covered by the single-film efficiency systematic of Section~\ref{sec:syst}.

The test-beam modules were irradiated with a target exposure of approximately 8 electrons/cm$^2$ and an estimated 1500 muons/cm$^2$, corresponding to a composition of $\sim$1\% EM showers and $\sim$99\% muons. The measured particle densities provide a cross-check of the reconstruction and identification efficiency, which factorizes as $\varepsilon_{total} = \varepsilon_{reco} \cdot \varepsilon_{id}$, where $\varepsilon_{id}$ denotes the Medium ID selection efficiency for signal and the background rejection rate for muons. The resulting overall efficiencies for signal are 58.9\% at 100\,GeV and 70.8\% at 200\,GeV. For muons, 28.8\% (100\,GeV) and 28.9\% (200\,GeV) pass pre-selection, of which 54.5\% and 55.5\% are successfully reconstructed; the identification chain then rejects 99.99\% and 99.94\% of these, leaving a negligible muon contamination in the final sample. After applying efficiency corrections, the average reconstructed EM densities are $3.7\pm0.4$\,cm$^{-2}$ (100\,GeV) and $7.9\pm0.5$\,cm$^{-2}$ (200\,GeV), with corresponding average muon densities of $1070\pm13$\,cm$^{-2}$ and $1533\pm14$\,cm$^{-2}$. The statistical uncertainty on each density is computed as $\sqrt{N}/(\varepsilon\cdot A)$, where $N$ is the raw count, $\varepsilon$ the corresponding efficiency, and $A$ the analyzed area, 40\,cm$^2$ for the 100\,GeV module and 48\,cm$^2$ for 200\,GeV. The 200\,GeV result is consistent with the nominal beam exposure target of $\sim$8\,electrons/cm$^2$. The 100\,GeV value is approximately a factor of two lower, which may reflect a combination of lower electron purity in the 100\,GeV beam during irradiation, the lower quality of the T1202 module, and the more challenging reconstruction at lower energy where smaller shower multiplicities reduce shower-to-background discrimination. The muon densities are affected by the same module quality differences, since
they count only muon tracks with a reconstructed segment on the first plate.

\section{Energy Reconstruction}
\label{sec:energy}

The total number of segments across all plates in the reconstructed shower, \ntotal, is used as the calorimetric energy estimator. This observable captures the full shower development and provides more robust energy estimation than localized estimators such as \nsegsum.

Since \ntotal depends directly on the single-film reconstruction efficiency, variations in efficiency systematically shift the observed shower multiplicity and consequently the reconstructed energy. To account for this, six independent MC samples are generated using GEANT4~\cite{Geant4} with single-film efficiencies of 75\%, 80\%, 85\%, 90\%, 95\%, and 100\%, each containing 3000 electrons with a flat energy distribution from 20 to 300\,GeV. For the 200\,GeV module, a close-pair inefficiency correction is applied to all MC samples prior to fitting. In dense shower cores at 200\,GeV, segment pairs separated by less than approximately 10\,\mum can be reconstructed as a single segment or fail reconstruction entirely; at 100\,GeV the sparser core renders it negligible. This effect is characterized by measuring reconstruction efficiency as a function of inter-segment distance and corrected for by applying a corresponding distance-dependent rejection probability to simulated segments. The correction reduces the mean total segment count in 200\,GeV MC by approximately 15\%, bringing the simulated shower multiplicity into agreement with the data and improving the linearity of the energy calibration at this operating point.

\begin{figure}[htbp]
\centering
\includegraphics[width=0.9\textwidth]{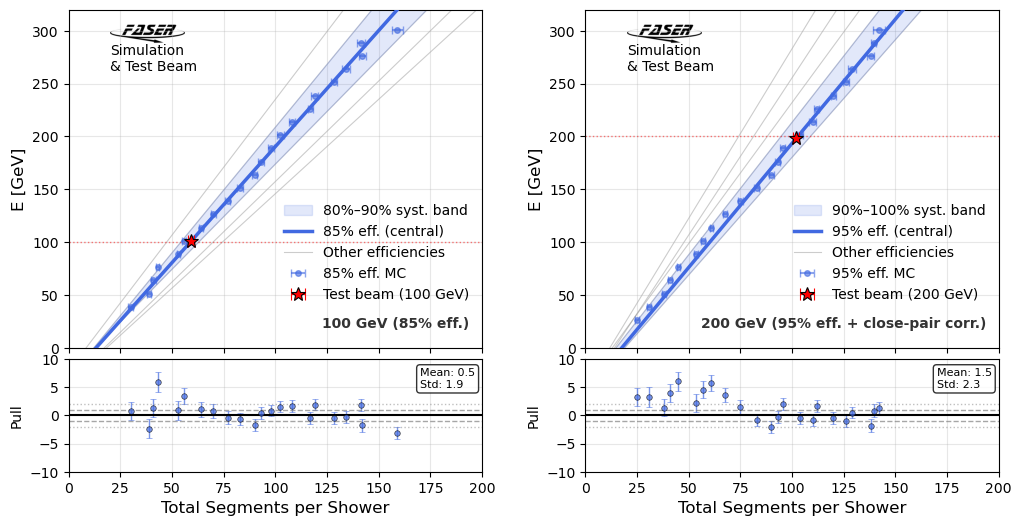}
\caption{Energy calibration using the total number of reconstructed segments as the estimator. Left: 100\,GeV operating point using 85\% single-film efficiency (thick blue line). Right: 200\,GeV operating point using 95\% efficiency with a close-pair correction applied to the MC. The systematic uncertainty from $\pm$5\% efficiency variation is shown as a blue band; thin grey lines show calibration curves at other efficiencies. Blue points show the median segment count per energy bin from MC, with horizontal error bars indicating the uncertainty on the mean ($\sigma_{\mathrm{mean}}=\sigma/\sqrt{N_\mathrm{bin}}$). Red stars show the test-beam data points with the same uncertainty definition. Lower panels show the pull $(E - E_{\mathrm{fit}})/\sigma_{\mathrm{mean}}$, with error bars $\sqrt{1 + (\sigma_{\mathrm{fit}}/\sigma_{\mathrm{mean}})^2}$, where $\sigma_{\mathrm{fit}}$ is the fit-parameter uncertainty propagated to energy via the calibration slope.}
\label{fig:calibration}
\end{figure}

For each efficiency scenario, a linear parametrization is fit:
\begin{equation}
N_{\mathrm{total}} = a(\varepsilon) \cdot E + b(\varepsilon)
\label{eq:calibration}
\end{equation}
where the slope $a$ and intercept $b$ depend on the single-film efficiency $\varepsilon$. Figure~\ref{fig:calibration} shows the energy calibration curves for both operating points.

When applying the calibration to test-beam data, the mean background contribution to \ntotal is subtracted from the measured shower segment count. This contribution is evaluated using randomly positioned cylinders (radius 100\,\mum, length 8\,cm, matching the shower reconstruction volume) placed throughout the detector volume, excluding regions within 100\,\mum of any reconstructed shower axis. From approximately 10{,}000 such cylinders per module, the mean contribution is 1.9 segments per shower at both 100\,GeV and 200\,GeV. The background is then subtracted before reading off the reconstructed energy for both modules.

\begin{figure}[htbp]
\centering
\includegraphics[width=0.8\textwidth]{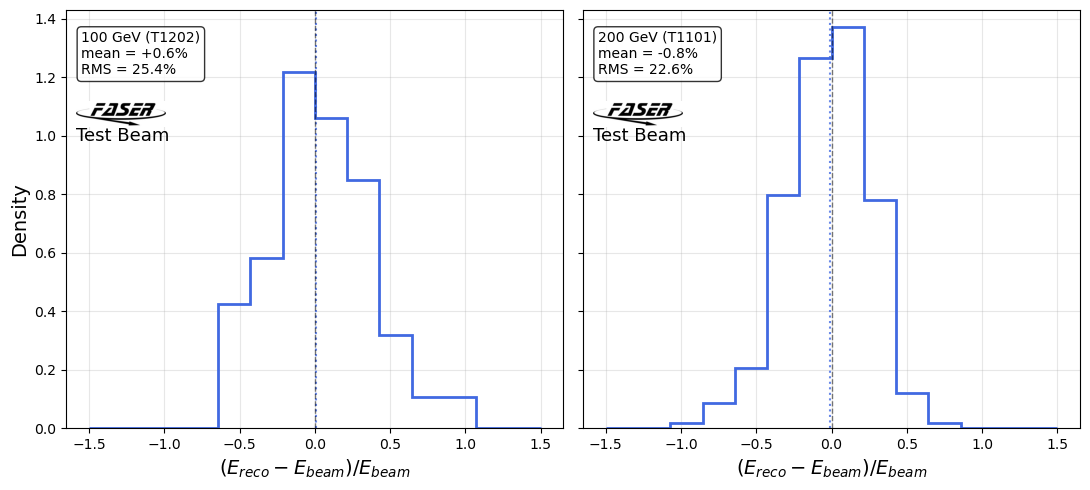}
\caption{Fractional residual distributions
$(E_\mathrm{reco} - E_\mathrm{beam})/E_\mathrm{beam}$ of the reconstructed
shower energies in test-beam data for the 100\,GeV (left) and 200\,GeV (right) modules. The mean and RMS of each distribution give the bias and resolution quoted in Table~\ref{tab:energy}.}
\label{fig:eresidual}
\end{figure}

The 85\% parametrization is applied to the 100\,GeV module (T1202) and the 95\% close-pair-corrected parametrization to the 200\,GeV module (T1101), consistent with the measured peak efficiencies. Table~\ref{tab:energy} summarizes the validation results. The reconstructed energies are in good agreement with the nominal beam energies, with relative biases of $+0.6\%$ at 100\,GeV and $-0.8\%$ at 200\,GeV. The energy resolution, defined as the RMS of the fractional residual distribution $(E_{\mathrm{reco}} - E_{\mathrm{beam}})/E_{\mathrm{beam}}$, is 25.4\% at 100\,GeV and 22.6\% at 200\,GeV, with the relative bias given by the mean
of the same distribution. The distributions are shown in
Figure~\ref{fig:eresidual}.

\begin{table}[htbp]
\centering
\caption{Energy reconstruction validation using test-beam data with the total segments estimator.}
\label{tab:energy}
{\setlength{\tabcolsep}{8pt}
\begin{tabular}{lccccc}
\hline
Beam energy & Efficiency & $\langle N_{\mathrm{total}} \rangle$ & $\langle E_{\mathrm{reco}} \rangle$ & Bias & Resolution \\
\hline
100\,GeV & 85\%                         & 59.1  & 100.6\,GeV & $+0.6\%$ & 25.4\% \\
200\,GeV & 95\% + close-pair corr.      & 102.1 & 198.4\,GeV & $-0.8\%$ & 22.6\% \\
\hline
\end{tabular}}
\end{table}

\section{Systematic Uncertainties}
\label{sec:syst}

The systematic uncertainties in this analysis are dominated by three sources: the beam energy spread, background contamination, and single-film efficiency fluctuations. 

\textbf{The beam energy spread} at the SPS H4 beamline is $\Delta p / p \approx1.4\%$~\cite{H4beamline}, which propagates directly to a $\sim$1.4\% uncertainty on the reconstructed energy. The beam has a transverse size of approximately 5\,mm in $X$ and 10\,mm in $Y$; this is small compared to the emulsion module dimensions and does not contribute a significant additional systematic.

\textbf{Background contamination} in \ntotal was measured using 10{,}000 random cylinders per module (Sec.~\ref{sec:energy}). The average contamination is found to be 1.9 segments per shower at both 100\,GeV and 200\,GeV. The statistical precision on these means is 0.06 segments at 100\,GeV and 0.08 segments at 200\,GeV, corresponding to a systematic uncertainty on the reconstructed energy of $\sim$0.1\% at both energies.

\textbf{Single-film efficiency} directly affects the reconstructed total segments and consequently the energy estimate. The central calibration uses peak efficiency values (85\% for the 100\,GeV module; 95\% for the 200\,GeV module), but efficiency varies across the detector volume, introducing a systematic uncertainty.

The uncertainty is quantified by evaluating energy reconstruction using calibration curves at $\pm$5\% efficiency relative to the central value. For the 100\,GeV module, the 80\% and 90\% standard MC parametrizations are used as bounds; for the 200\,GeV module, the 90\% and 100\% close-pair-corrected parametrizations are used. The resulting reconstructed energies are:

For the 100\,GeV module with $\langle N_{\mathrm{total}} \rangle = 59.1$:
\begin{itemize}
    \item Central (85\% efficiency): $E_{\mathrm{reco}} = 100.6$\,GeV
    \item 80\% efficiency bound: $E_{\mathrm{reco}} = 111.5$\,GeV ($+10.8\%$)
    \item 90\% efficiency bound: $E_{\mathrm{reco}} = 92.4$\,GeV ($-8.1\%$)
\end{itemize}

For the 200\,GeV module with $\langle N_{\mathrm{total}} \rangle = 102.1$:
\begin{itemize}
    \item Central (95\% efficiency + close-pair corr.): $E_{\mathrm{reco}} = 198.4$\,GeV
    \item 90\% efficiency + close-pair corr. bound: $E_{\mathrm{reco}} = 218.6$\,GeV ($+10.2\%$)
    \item 100\% efficiency + close-pair corr. bound: $E_{\mathrm{reco}} = 184.8$\,GeV ($-6.8\%$)
\end{itemize}

The asymmetry of the uncertainties reflects the non-linear dependence of the calibration slope on efficiency: since $a(\varepsilon)$ varies with efficiency (Figure~\ref{fig:calibration}), the same measured \ntotal maps to different reconstructed energies depending on the assumed efficiency. The larger uncertainty at 100\,GeV reflects the steeper efficiency dependence at lower energies, where smaller absolute segment counts amplify the fractional impact of efficiency variations.

The single-film efficiency variation also propagates to the energy resolution. Changing the assumed efficiency replaces the calibration slope $a$ by $a'$, which rescales every reconstructed energy by a common factor $k = a/a'$, while the intercept shifts the distribution without changing its width. This gives resolutions of $25.4^{+2.1}_{-2.4}\%$ at 100\,GeV and $22.6^{+1.5}_{-1.2}\%$ at 200\,GeV.

Table~\ref{tab:systematics} summarizes the systematic uncertainties on EM shower energy reconstruction. Single-film efficiency variations dominate at both energies. Background contamination remains negligible under test-beam conditions.

\begin{table}[htbp]
\centering
\caption{Summary of systematic uncertainties on electromagnetic shower energy reconstruction in test-beam data.}
\label{tab:systematics}
\begin{tabular}{lcc}
\hline
Source & 100\,GeV & 200\,GeV \\
\hline
Beam energy spread      & $\pm1.4$\%                      & $\pm1.4$\%                      \\
Background contamination & $\sim$0.1\%                     & $\sim$0.1\%                      \\
Single-film efficiency  & $^{+10.8}_{-8.1}\%$      & $^{+10.2}_{-6.8}\%$      \\
\hline
\textbf{Total}          & $\mathbf{^{+10.9}_{-8.2}\%}$ & $\mathbf{^{+10.3}_{-6.9}\%}$ \\
\hline
\end{tabular}
\end{table}

\section{Conclusion}
\label{sec:conclusion}

We have presented and validated methods for electromagnetic shower reconstruction and identification in the \fasernu emulsion detector using 100\,GeV and 200\,GeV electron test-beam data from the CERN SPS H4 beamline. The reconstruction algorithm combines adaptive DBSCAN clustering, sequential centroid validation with a 60\mum deviation threshold, and Huber robust fitting to determine shower axes without energy-dependent parameter tuning.

A multi-level identification chain comprising track pre-selection, cut-based Loose ID, and BDT-based Medium ID achieves combined background rejection rates of 99.99\% at 100\,GeV and 99.94\% at 200\,GeV, with signal efficiencies of 58.9\% and 70.8\% respectively, evaluated from MC simulations. The BDT classifier uses ten shower shape variables: six longitudinal, three transverse and clustering, and one angular, achieving AUC = 0.997. The ratio-based and angular variables are designed to remain stable across the broad \fasernu energy range, enabling a single trained classifier to be applied at 50\,GeV and above without retuning.

Energy reconstruction using the total reconstructed segments yields relative biases of $+0.6\%$ at 100\,GeV and $-0.8\%$ at 200\,GeV, with resolutions of $25.4^{+2.1}_{-2.4}\%$ and $22.6^{+1.5}_{-1.2}\%$. Systematic uncertainties on the reconstructed energy are dominated by single-film efficiency variations, contributing $^{+10.8}_{-8.1}\%$ and $^{+10.2}_{-6.8}\%$ at the two energies, with totals of $^{+10.9}_{-8.2}\%$ and $^{+10.3}_{-6.9}\%$. Background contamination remains below 0.1\% under test-beam conditions. 

Once validated in the high-background LHC environment with appropriate beam track masking, this method will be deployed for \fasernu neutrino interaction analyses measuring electron neutrino charged-current interaction cross-sections and studying neutrino flavour composition at TeV energies.

\section*{Acknowledgments}
\label{sec:Acknowledgments}
We thank the CERN EP department for the refurbishment of the CERN dark room, used for the preparation and development of the FASER$\nu$ emulsion. We thank Saya Yamamoto for supporting the preparation of emulsion films. We thank the CERN beam group, especially Nikolaos Charitonidis, for excellent beam operation at the SPS-H4 beamline.

This work was supported in part by Heising-Simons Foundation Grant Nos.~2019-1179 and 2020-1840, Simons Foundation Grant No.~623683, JSPS KAKENHI Grant Nos.~19H01909, 22H01233, 20K23373, 23H00103, 20H01919, and 21H00082, the joint research program of the Institute of Materials and Systems for Sustainability, ERC Consolidator Grant No.~101002690, DFG grant~SCHO~1527/13-1, Royal Society Grant No.~URF$\backslash$R1$\backslash$201519, UK Science and Technology Funding Councils Grant No.~ST/ T505870/1, the National Natural Science Foundation of China, Tsinghua University Initiative Scientific Research Program, and the Swiss National Science Foundation.

\appendix
\section{Data Quality of the Test-Beam Modules}
\label{app:dataquality}
This appendix presents the film-by-film data quality measurements underlying the summary values in Table~\ref{tab:dataquality}. All quantities are evaluated using the methods of Ref.~\cite{FASER:reco}.
 
\begin{figure}[htbp]
\centering
\includegraphics[width=0.45\textwidth]{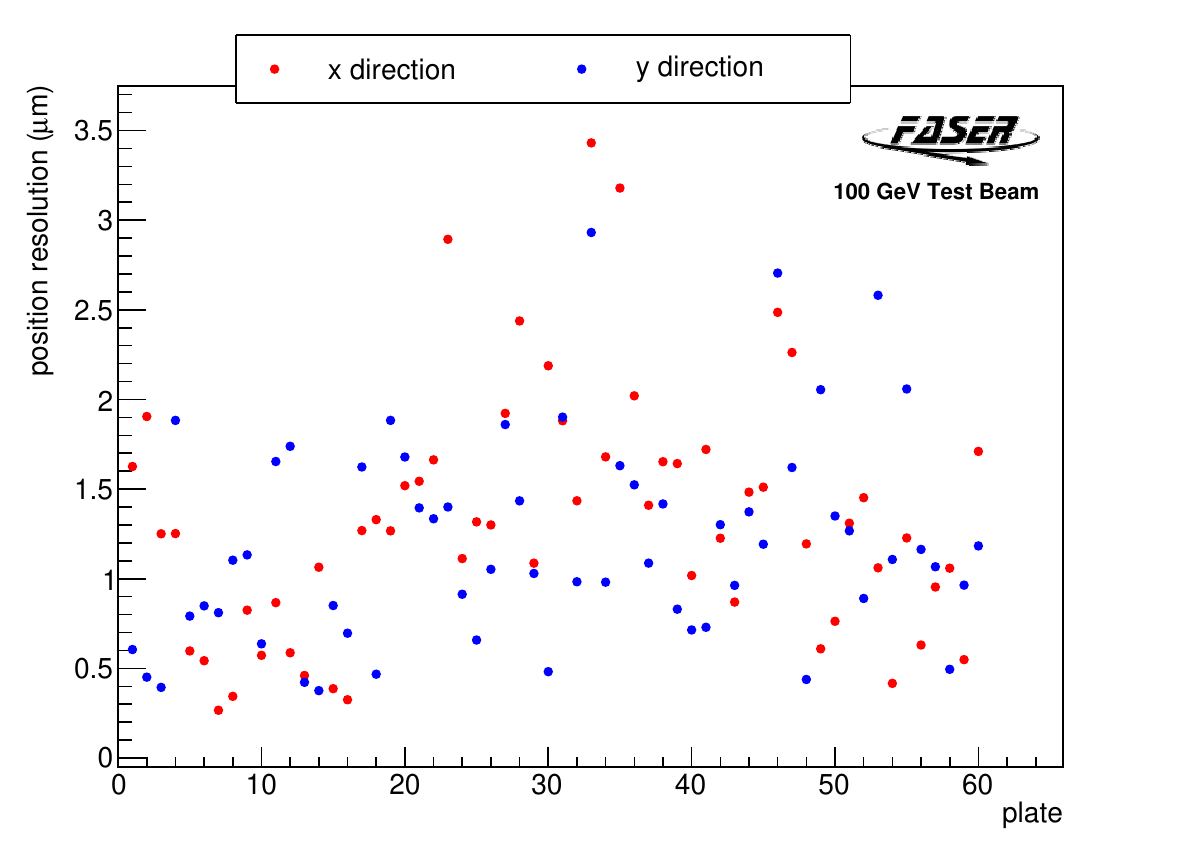}
\includegraphics[width=0.45\textwidth]{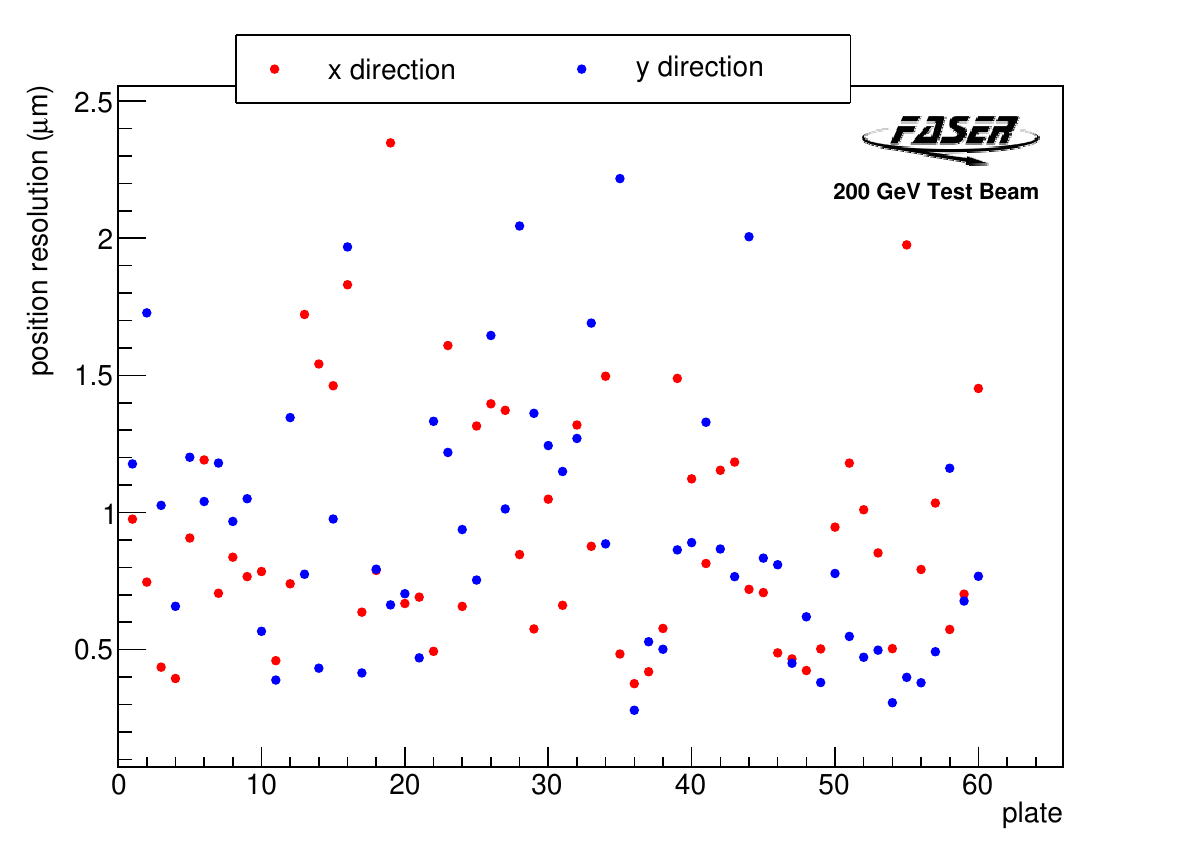}
\caption{Position resolution versus plate number in the $x$ and $y$ projections for the 100\,GeV (left) and 200\,GeV (right) modules.}
\label{fig:posresvsplate}
\end{figure}

The position resolution is evaluated with a track-extrapolation method: for
each film, a linear trajectory is reconstructed from the base-tracks on the
two films immediately upstream and two immediately downstream, and the
residual is the deviation between the measured and extrapolated positions.
Figure~\ref{fig:posresvsplate} shows the resolution as a function of plate
number for both modules. Most plates exhibit resolutions of 0.5--1.5\,$\mu$m, with several films degraded up to $\sim$3\,$\mu$m due to variations in film alignment quality, emulsion distortions, and scanning conditions. 

\begin{figure}[htbp]
\centering
\includegraphics[width=0.45\textwidth]{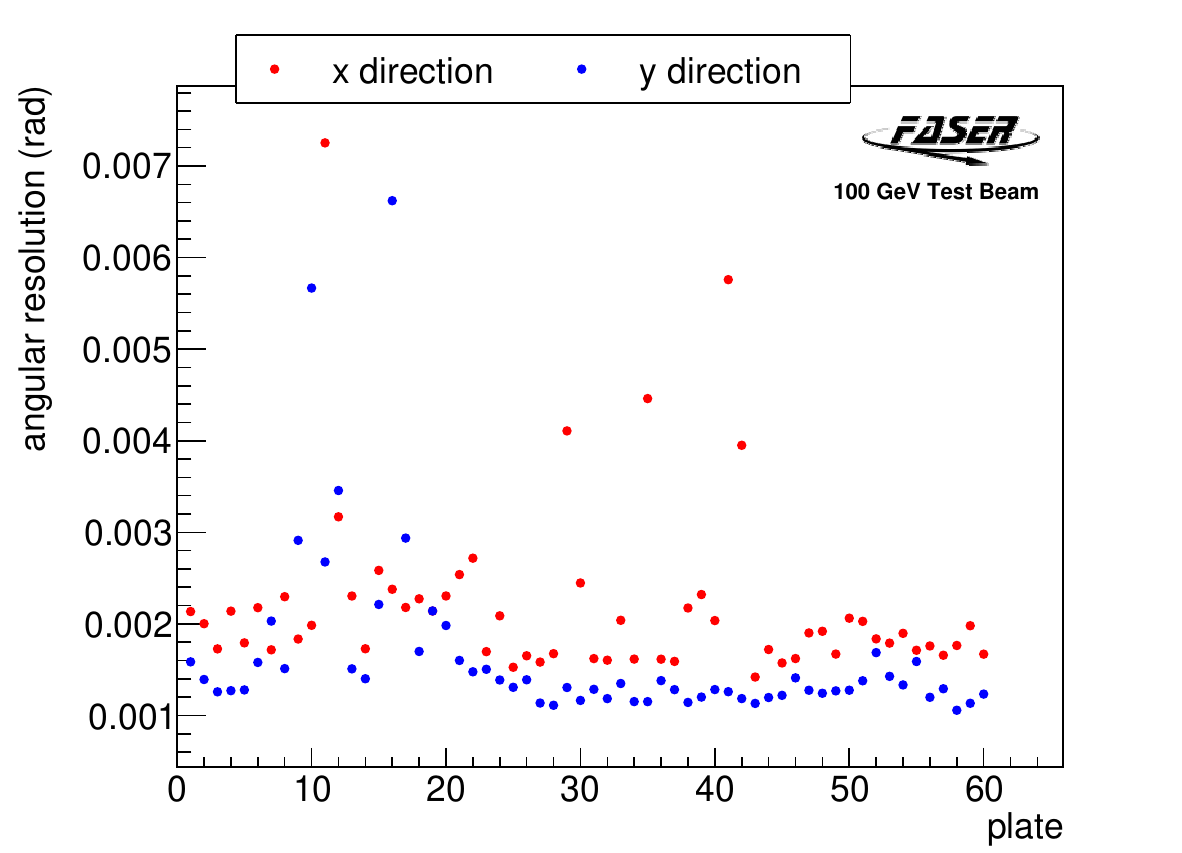}
\includegraphics[width=0.45\textwidth]{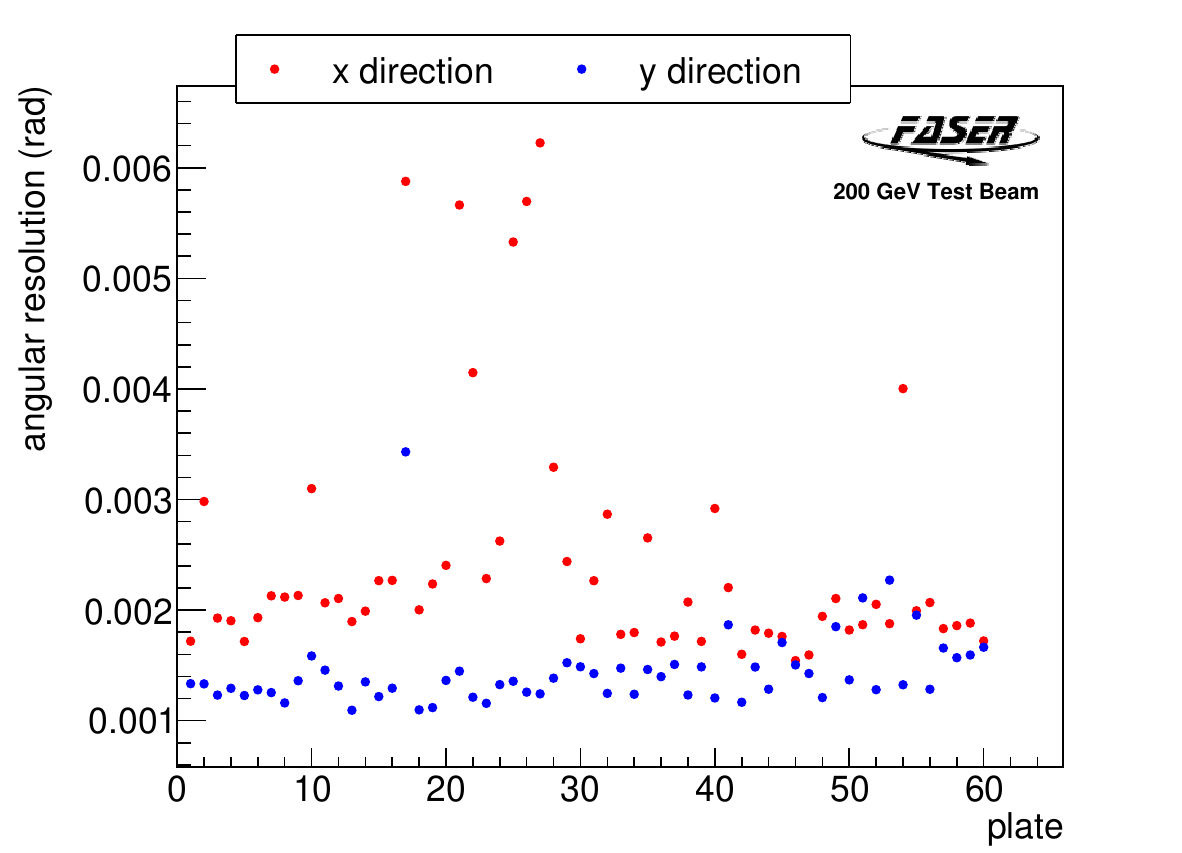}
\caption{Angular resolution versus plate number in the $x$ and $y$
projections for the 100\,GeV (left) and 200\,GeV (right) modules.}
\label{fig:angresvsplate}
\end{figure}

The angular resolution is evaluated with a segment-difference method: the
track direction is fit independently from two separated groups of ten
consecutive segments, and the discrepancy between the two estimates
characterizes the combined effect of measurement uncertainty and residual
alignment (Figure~\ref{fig:angresvsplate}). The resolution is systematically larger in the horizontal projection in both modules, reflecting the elliptical angular distribution of the test beam, which has a larger divergence in $\tan\theta_x$ than in $\tan\theta_y$.

For each target film $i$, the single-film efficiency is defined using the two films upstream and two downstream as
\begin{equation}
\varepsilon_i =
\frac{N_\mathrm{tracks~with~segments~on~films}~(i{-}2,\,i{-}1,\,i,\,i{+}1,\,i{+}2)}
     {N_\mathrm{tracks~with~segments~on~films}~(i{-}2,\,i{-}1,\,i{+}1,\,i{+}2)}\,.
\label{eq:sfe}
\end{equation}
The four-film bracketing constrains the trajectory when testing for a segment on the target film; the segment search window is large compared to the position resolution, so the efficiency measurement is insensitive to the resolution variations discussed above.
 
\begin{figure}[htbp]
\centering
\includegraphics[width=0.45\textwidth]{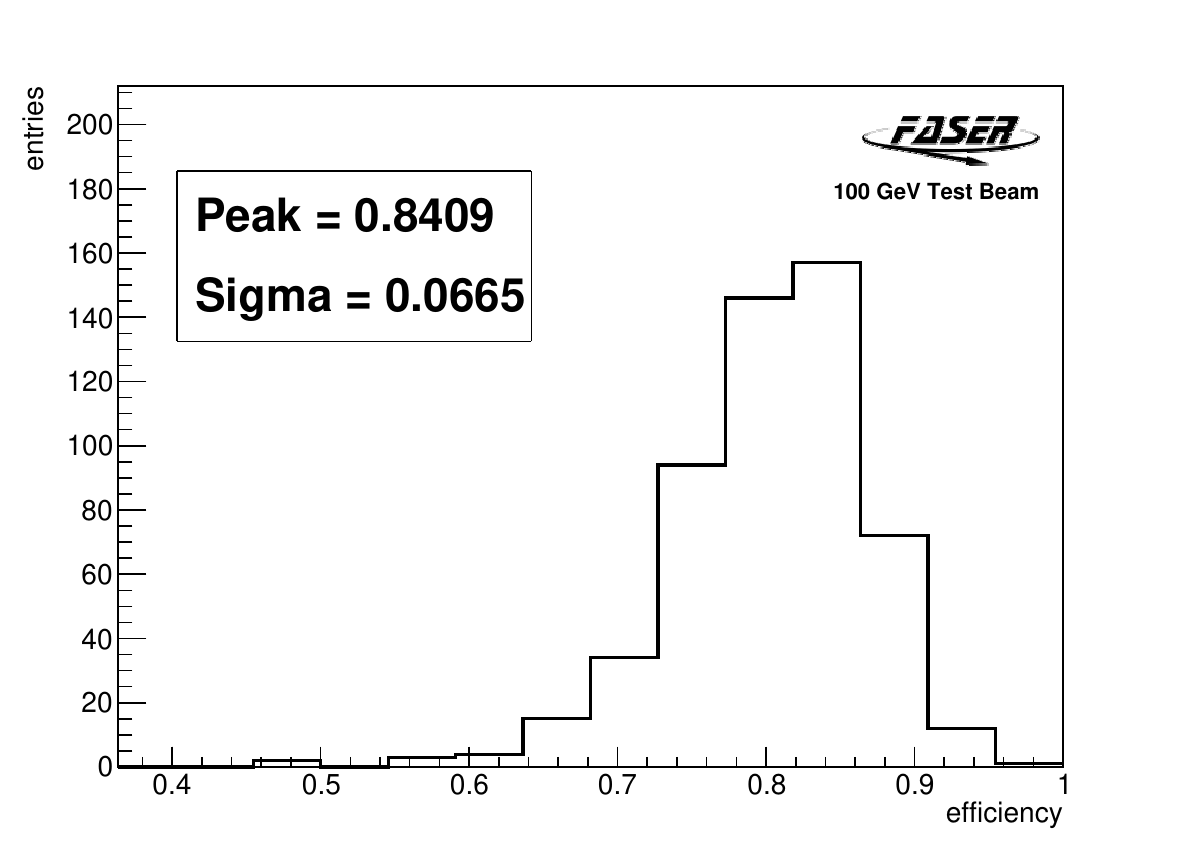}
\includegraphics[width=0.45\textwidth]{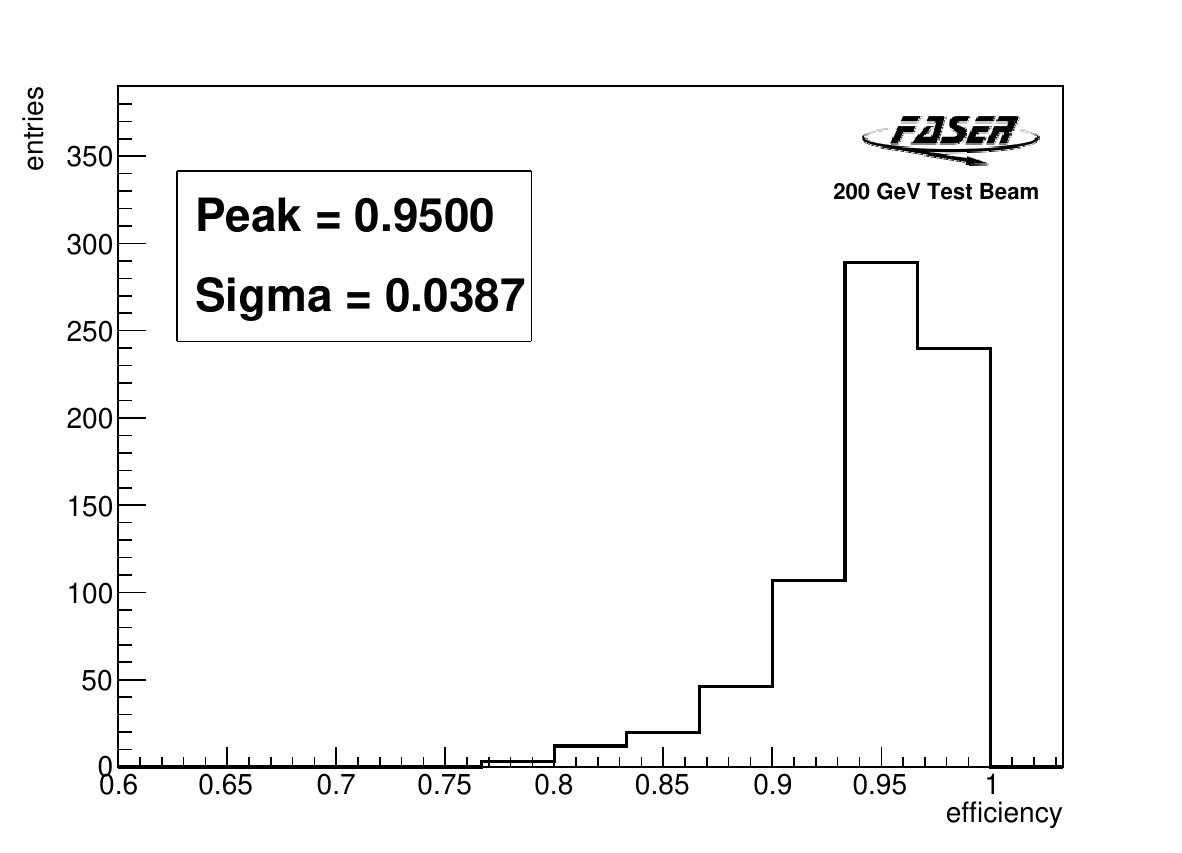}
\caption{Film-by-film single-film efficiency distributions for the 100\,GeV
(left) and 200\,GeV (right) modules, aggregated over all films. The 
peak and standard deviation of each distribution are indicated.}
\label{fig:effdata}
\end{figure}

Figure~\ref{fig:effdata} shows the resulting distributions. The 100\,GeV module peaks at $\sim85$\% with a spread of $\sigma = 6.7\%$; the 200\,GeV module peaks at $\sim95$\% with $\sigma = 3.9\%$. The 100\,GeV module thus shows both a lower average efficiency and a larger film-to-film variation; 
these measured distributions motivate the $\pm$5\% efficiency variation used for the systematic uncertainty in Section~\ref{sec:syst}.
 
\section{Energy Dependence of the Identification Variables}
\label{app:idvars}
 
\begin{figure}[htbp]
\centering
\includegraphics[width=\textwidth]{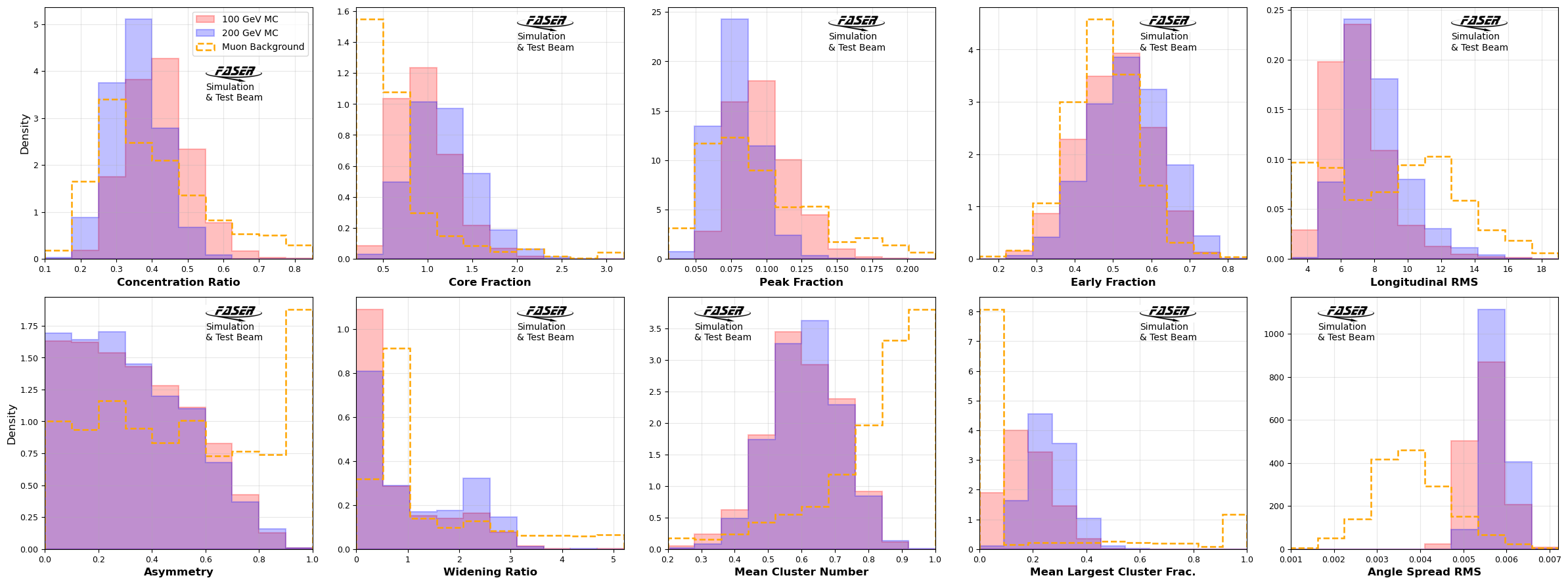}
\caption{Distributions of the ten BDT input variables for electron MC at
100\,GeV (red) and 200\,GeV (blue), together with the muon background from
test-beam data identified by visual inspection (orange dashed). Only the
100\,GeV MC and the background are used for classifier training.}
\label{fig:bdtvarscomp}
\end{figure}

Figure~\ref{fig:bdtvarscomp} compares the ten BDT input variables between the 100\,GeV and 200\,GeV electron MC samples, together with the muon background sample of Figure~\ref{fig:features}. Only the 100\,GeV MC sample is used for classifier training; the 200\,GeV sample is shown here to test how the variables behave at an energy the classifier has not seen. The signal distributions have similar shapes at the two energies, with modest shifts reflecting the longer and denser showers at higher energy. 
 
\begin{figure}[htbp]
\centering
\includegraphics[width=\textwidth]{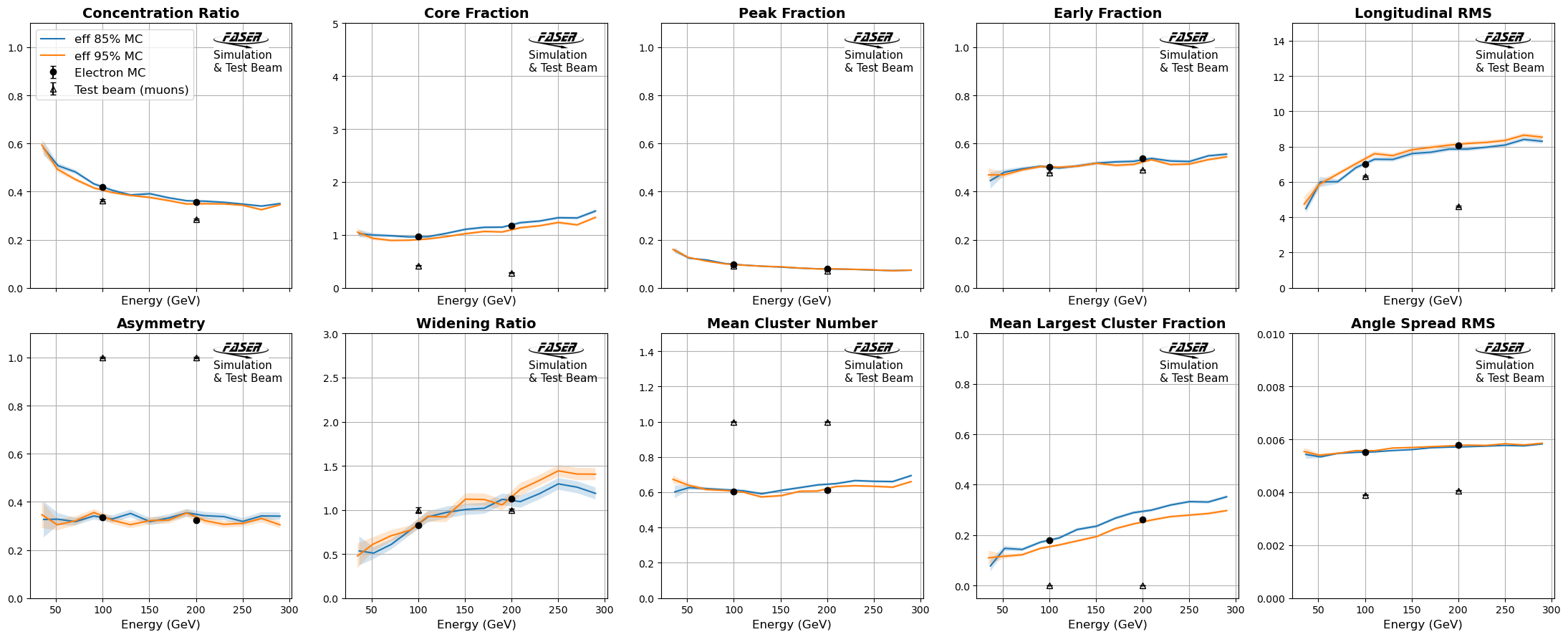}
\caption{The ten BDT input variables as a function of electron energy,
evaluated on flat-energy-spectrum electron MC for single-film efficiencies of 85\% and 95\% (the latter with the close-pair correction). Lines show the mean and bands the uncertainty on the mean. Black circles show the electron MC at the two test-beam energies, 100\,GeV and 200\,GeV. Triangles show the muon-dominated test-beam sample before identification; their offset from the signal curves reflects the separation between signal and background.}
\label{fig:varsvsenergy}
\end{figure}

Figure~\ref{fig:varsvsenergy} shows the same variables as a continuous
function of energy, evaluated on flat-energy-spectrum electron MC samples generated with single-film efficiencies of 85\% and 95\%. Above approximately 50\,GeV, the relevant range for $\nu_e$ CC electrons in \fasernu, the variables vary only weakly and smoothly with energy; the angular spread RMS $\sigma_\theta$ is stable across the full range. 
 
\bibliographystyle{utphys}
\bibliography{references}

\end{document}